\documentclass[preprint,12pt]{elsarticle}

\usepackage{amssymb}
\usepackage{amsmath}
\usepackage{todonotes}
\usepackage{hyperref}
\usepackage{comment}
\usepackage{booktabs}
\usepackage{tabularx}
\usepackage{pdflscape}
\usepackage{graphicx}
\usepackage{textcomp}
\usepackage{multicol}
\usepackage{outlines}
\usepackage{multirow}
\usepackage{lscape}
\usepackage{comment}
\usepackage[table,xcdraw,dvipsnames]{xcolor}
\usepackage{float}
\usepackage{enumitem}
\usepackage{ctable}
\usepackage{caption}
\usepackage[many]{tcolorbox}
\usepackage{nameref}
\usepackage{hyperref}
\usepackage{makecell}
\usepackage{hhline}
\usepackage{ctable}
\usepackage{array}
\usepackage{comment}
\usepackage{longtable}
\usepackage[T1]{fontenc}
\usepackage{soul}
\usepackage{tabularray}
\UseTblrLibrary{siunitx}

\newtcolorbox{colorb}{
enhanced,
boxrule=0pt,frame hidden,
borderline west={2pt}{0pt}{green!50!black},
colback=green!05!white,
sharp corners
}
\newtcolorbox{colora}{
enhanced,
boxrule=0pt,frame hidden,
borderline west={2pt}{0pt}{gray!50!black},
colback=gray!05!white,
sharp corners
}

\journal{Information and Software Technology}

\begin{document}

\begin{frontmatter}

\title{Privacy by Design: Aligning GDPR and Software Engineering Specifications with a Requirements Engineering Approach}

\author[label1,label2]{Oleksandr Kosenkov}
\author[label1]{Ehsan Zabardast}
\author[label1]{Davide Fucci}
\author[label1,label2]{Daniel Mendez}
\author[label1]{Michael Unterkalmsteiner}

\affiliation[label1]{organization={Blekinge Institute of Technology}, addressline={Valhallavägen 10}, city={Karlskrona},             postcode={371 79}, country={Sweden}}
\affiliation[label2]{organization={fortiss GmbH},
            addressline={Guerickestr. 25}, 
            city={Munich},
            postcode={80805}, 
            country={Germany}}

\begin{abstract}
\indent\textit{Context:} Consistent requirements and system specifications are essential for the compliance of software systems towards the General Data Protection Regulation (GDPR). Both artefacts need to be ``grounded'' in the original text and conjointly assure the achievement of privacy by design (PbD).
\textit{Objectives:}
There is little understanding of the perspectives of practitioners on specification objectives and goals to address PbD. Existing approaches to GDPR and PbD do not account for the complex intersection between problem and solution space expressed in GDPR. In this study we explore the demand for conjoint requirements and system specification for PbD and suggest an initial version of an approach to address this demand.
\textit{Methods:} We reviewed existing secondary and related primary studies on GDPR compliance and conducted interviews with practitioners to (1) investigate the state-of-practice in requirements and system specifications for GDPR compliance and (2) understand the underlying specification objectives and goals (e.g., traceability). We developed and evaluated an initial version of an approach for requirements and systems specification for PbD, and evaluated it against the specification objectives.
\textit{Results:} The relationship between problem and solution space, as expressed in GDPR, is instrumental in supporting PbD. We demonstrate how our approach, based on the modeling GDPR content with original legal concepts, contributes to specification objectives of capturing legal knowledge, supporting specification transparency for roles involved, and traceability.
\textit{Conclusion:} In addition to assuring traceability, GDPR demands need to be addressed throughout different levels of abstraction in the engineering lifecycle to achieve PbD. Legal knowledge specified in the GDPR text should be captured in specifications to address the demands of different stakeholders and ensure compliance. While our results confirm the suitability of our approach to address practical needs, we also revealed specific needs for the future effective operationalization of our suggested approach.
\end{abstract}


\end{frontmatter}

\section{Introduction}\label{sec:intro}
Regulations are an important source of requirements and constraints for software systems. Recent regulations often specifically address digital technologies and their social impact. In particular, the General Data Protection Regulation (GDPR) has a profound impact on software engineering (SE) research and practice~\cite{leite2022impact}. One principle emerging across multiple recent regulations is compliance by design, or ``privacy by design'' (PbD) in GDPR.

PbD requires compliance controls to be built into software design and, hence, it has an impact on the software design and architecture (SDA) phase of software development life cycle (SDLC). However, GDPR must be effectively processed in the requirements engineering (RE) phase of the SDLC to ensure that the GDPR demands are consistently communicated and considered in SDA. There are at least three reasons to consider the intersection between RE and SDA. First, although regulatory compliance has a de-facto impact on multiple SDLC phases~\cite{kempe2021perspectives}, it remains challenging to seamlessly integrate compliance throughout the SDLC in practice~\cite{leite2022impact}, ensuring the consistency of engineering artefacts. This is important to yield a reproducible and consistent evidence for compliance implementation throughout the SDLC. Second, GDPR, like other regulations, addresses heterogeneous aspects of software (software quality and user behavior~\cite{breaux2008analyzing}).
As a result, requirements derived from regulations are often concerned both with the \textit{``What shall be developed?''} and the \textit{``How it shall be developed?'')}~\cite{marques2019analysis}, rendering relationships between requirements and downstream SDLC phases essential.
Third, interpretation of intentionally abstract regulations into system-specific demands requires legal knowledge and interpretation rules~\cite{bobkowska2010efficient}. This hinders the transition and traceability between the RE and SDA phases of SDLC.

Specifications (e.g., requirements specification) capture and communicate the required information so that a system is implemented as expected throughout the SDLC~\cite{alagar2011role}. While implementing PbD, it remains unclear what objectives and goals should be achieved with requirements and system specification methods conjointly. Furthermore, empirical studies allowing the community to address this challenge are lacking. As a result, we lack methods that facilitate the engineering of requirements and system specifications in an integrated and consistent manner for PbD.

We conducted a study to address the knowledge gap on effective requirements and system (R\&S) specification for PbD. Our aim is to (1) identify the state of research and practice in conjoint R\&S specification for GDPR compliance, (2) introduce an integrated R\&S specification approach, and (3) evaluate the usefulness of our approach towards the specification objectives.

We conducted a literature review to identify existing approaches to R\&S specification for PbD and synthesize the R\&S specification objectives. Accordingly, we created a list of requirements for our R\&S specification approach. Next, we conducted interviews with practitioners involved in R\&S specification, focusing on the specification objectives to identify the existing state of practice. Finally, we evaluated our approach through screening and experiment methodologies involving practitioners~\cite{kitchenham1996desmet, wohlin2012experimentation}.

The contributions of our study are:
\begin{enumerate}[nosep]
    \item The identification of five main R\&S specification objectives for PbD that characterize the required R\&S specification methods. 
    \item An overview of R\&S specification goals that practitioners strive to achieve with the application of R\&S specification methods.
    \item A systematic R\&S specifications' content modeling approach for PbD and its initial evaluation.
\end{enumerate}

Our study contributes to the existing corpus of knowledge on PbD and GDPR compliance in two principal ways. Firstly, we explicitly address the intersection between requirements and system specifications for PbD. We do so by identifying corresponding specification objectives and goals, and, building upon this foundation, by proposing a method addressing this intersection. Our findings facilitate the systematization of the integrated requirements and system specification for PbD by suggesting the priorities of specification goals that requirements engineering methods should fulfill. Secondly, our proposed requirements engineering approach for R\&S specifications' content modeling systematically captures the legal domain knowledge and viewpoint for subsequent transfer into software specifications. Such systematic processing---in contrast with the ad hoc requirements derivation and ontology construction applied in prior work---is essential for addressing future changes in GDPR, processing auxiliary regulatory sources (e.g., guidelines). Furthermore, the approach has the potential to enhance the verifiability of PbD implementation by making the interpretation of GDPR transparent to both legal and engineering roles, thereby facilitating communication between them.

The remainder of this paper is organized as follows. In Section~\ref{sec:relatedWork}, we define the key concepts central to this study. Section~\ref{sec:relatedWork} reviews existing related studies. Section~\ref{sec:methodology} presents the research methodology employed. In Section~\ref{sec:results}, we detail our findings and introduce our proposed R\&S specification approach in Section~\ref{sec:resultsSynthesis}. Section~\ref{sec:discussion} discusses the main insights derived from the results, followed by a discussion of the threats to validity in Section~\ref{sec:threatsValidity}. Section~\ref{sec:conclusion} concludes the paper and outlines directions for future work. The open dataset supporting this study is publicly available on \href{https://zenodo.org/records/15565632}{Zenodo (DOI: 10.5281/zenodo.15565632)}.

\section{Background}\label{sec:background}
This section introduces the main terms and concepts used in this study.

\paragraph{Regulatory Compliance in Software Engineering}\label{sec:compliance}
We define \textit{regulatory compliance} as a state of verifiable conformance of software systems to requirements emerging from regulations~\cite{kosenkov2024systematic}. GDPR contains general norms that need to be interpreted into system and project-specific requirements and system specifications. In the context of this study, we suggest that verifiable conformance entails specifying such interpretation in a form that (1) allows its review by both engineers and legal experts and that (2) supports other verification evidence (e.g., automated testing results).

GDPR is the European personal data protection regulation enacted in 2016 and enforced from 2018. Article 25 of GDPR introduces the principle of ``personal data protection by design''(DPbD), requiring that privacy controls are designed and embedded into the system at design-time. The principle of DPbD resembles the principle of privacy by design (PbD), meaning ``embedding privacy into information technologies, as a core functionality, from the outset''~\cite{cavoukian2012operationalizing}. From an engineering perspective, both personal data protection by design and privacy by design emphasize an approach requiring controls to be effectively implemented in the software architecture phase of SDLC. We define \textit{privacy by design (PbD)} as the specification of requirements and early software design and architecture in response to GDPR norms, facilitating demonstrable and verifiable compliance. The practice and research, PbD sometimes includes additional activities (data protection impact assessment, threat modeling~\cite{herwanto2024towards, notario2015pripare}). GDPR contains only some basic requirements and procedure descriptions for such activities, while detailed demands towards them are covered in additional regulatory sources like guidelines. These procedural demands are not directly applicable to system or requirements specifications; rather, they pertain to the overarching conduct of processes that may subsequently prompt modifications to existing specifications. Accordingly, this study does not consider such activities due to their indirect relevance to the substantive content of requirements and system specifications. However, we plan to address such additional activities in our future work. PbD is also sometime considered in a broad sense as including, along with legal concerns, ethical, and personal privacy demands ~\cite{gurses2011engineering}. Our definition of PbD is targeting the core of PbD---the implementation of GDPR norms in software R\&S specification. We use the term PbD interchangeably with the terms personal data protection by design, GDPR compliance by design, and R\&S specification for GDPR compliance.

\paragraph{Requirements, and System Specification}\label{secLspecification}
Specifying engineering-relevant information (e.g., requirements) in SE is essential both in non-agile and agile SDLC models.
How exactly it is specified with respect to chosen notations is subject to various factors, among them the chosen process model. In our study, we take a process-agnostic view.

Academic literature emphasizes the importance of the separation of problem space and solution space specification. In practice, problem and solution spaces are closely intertwined and delimitation between them is vague~\cite{de2009similarity}, unobtainable or undesirable~\cite{partridge1995specification}.
In this study, we define \textit{requirements specification} as the specification of the problem space (What to develop?) addressed in SE. We define \textit{system specification} as specification of solution space (How to develop?). We follow a broad approach to system specification and consider that it embraces any form of early software design and architecture (e.g., architectural prescriptions and constraints). We use the term \textit{system component} also to denote problem space concerns imposing structure or limitation to the solution space that can be attributed, or lead to the creation of a separate software system component.

We introduce further terms related to R\&S specifications. \textit{Specification objective (SO)} is an immediate outcome achieved through the application of a method (e.g., specifications' consistency). Specification objective characterizes specification method and answers the question ``What should be achieved immediately by applying this method?''. On the other hand, \textit{specification goal} is a final purpose for which an outcome of a specification method is used and which drives the specification process. For example, the specification objective of consistency contributes to the goal of managing risks.

\section{Related work}\label{sec:relatedWork}
In this section, we provide an overview of four main clusters of related work which are as follows (1) studies on GDPR compliance in software engineering, (2) PbD in software engineering, (3) requirements and system specification for GDPR compliance, (4) modeling of regulations for GDPR compliance. We discuss each of these clusters next. Table~\ref{tab:relatedWork} compares the scope of the selected related studies to the scope of our study.

\begin{table*}[ht]
\tiny
\renewcommand\theadfont{\tiny}
\setlength{\tabcolsep}{3.75pt}
\centering
\rowcolors{3}{gray!15}{white}
\begin{tabular}{lccccccccccc}
 & \cite{diamantopoulou2021integrating} & \cite{bonatti2020real} & \cite{cejas2023nlp} & \cite{sion2020dpmf} & \cite{artemiou2019pdp4e,artemiou2019pdp4e2} & \cite{hjerppe2019general} & \cite{shah2019analyzing} & \cite{tomashchuk2020operationalization} & \cite{istvan2021software} & \cite{stefanova2021privacy} & \textbf{This study} \\
\textbf{Compliance scope} &  &  &  &  &  &  &  &  &  &  &  \\
- business processes & x &  &  &  &  &  &  &  &  &  &  \\
- privacy policies / agreements &  &  & x &  &  &  &  &  &  &  &  \\
- data control mechanisms &  & x &  &  &  &  &  &  &  &  &  \\
- software system (development) &  &  &  & x & x & x & x & x & x & x & x \\
\textbf{Specification / modeling scope} &  &  &  &  &  &  &  &  &  &  &  \\
- requirements & x & x & x &  & x &  &  &  &  &  &  \\
- system (architecture) &  &  &  &  & x &  &  &  &  &  &  \\
- requirements and system conjointly &  &  &  & x &  & x & x & x & x & x & x \\
\textbf{Specification / modeling purpose} &  &  &  &  &  &  &  &  &  &  &  \\
- legal activities & x &  & x & x &  &  &  &  &  &  & x \\
- business process management & x &  &  &  &  &  &  &  &  &  &  \\
- engineering activities &  & x & x &  & x & x & x & x & x & x & x \\
\textbf{PbD explicitly addressed} & x &  &  & x & x & x &  &  &  &  & x \\
\textbf{GDPR coverage} &  &  &  &  &  &  &  &  &  &  &  \\
- complete &  &  & x &  & x & x &  &  &  &  & x \\
- incomplete & x & x &  & x &  &  & x & x & x & x &  \\
\textbf{Spec. / model. based on GDPR} &  &  &  &  &  &  &  &  &  &  &  \\
- systematic &  & x & x &  &  &  &  &  &  &  & x \\
- ad hoc & x &  &  & x & x & x & x & x & x & x &  \\
\textbf{Empirical evaluation} &  &  & x &  & x &  & x &  &  &  & x
\end{tabular}
\caption{Table comparing the scope of this study to selected prior work. Covered aspects are marked with ``x''.}
\label{tab:relatedWork}
\end{table*}

\paragraph{GDPR Compliance in Software Engineering}\label{sec:complianceSe}
Since GDPR was enacted, multiple research studies and projects were completed to address GDPR compliance~\cite{de2020protecting}.
The majority of the projects and related studies in terms of the compliance scope were focused on GDPR compliance of business processes~\cite{diamantopoulou2021integrating},  standalone solutions for GDPR compliance (e.g.,~\cite{de2020protecting}) , development of platforms for GDPR compliance (e.g.,~\cite{ferrara2018static,piras2019defend}), data control mechanisms (e.g., data usage policies~\cite{bonatti2020real}), or compliance of privacy policies or data processing agreements (e.g.,~\cite{cejas2023nlp}), rather than incorporation of GDPR compliance throughout the SDLC phases. One example of such study is the Data Protection Modeling Framework~\cite{sion2020dpmf, sion2023demonstration}. The framework aims to capture the personal data processing architecture across multiple software systems to assist in legal assessments, rather than to support SE activities.

One initiative specifically focused on GDPR compliance in SE was PDP4E project~\cite{martin2018methods}. The specification scope in this project included both requirements and system, however, requirements engineering and software design SDLC phases were considered in different work packages, and no explicit connection between R\&S specifications was established.

Recent studies recognized that SDA could benefit from considering RE research~\cite{caiza2019reusable,sion2019architectural}. Recent RE studies also recognize that RE would benefit from considering subsequent SDLC phases in the context of regulatory compliance implementation~\cite{kempe2021regulatory}.
However, there are only a few studies considering R\&S specification conjointly. We discuss these studies next.

\paragraph{Privacy by Design in Software Engineering}\label{sec:pbdSe}
Studies in this research track often focus on RE (e.g.,~\cite{alshammari2017towards, lentzsch2017integrating}) or SDA perspective (e.g.,~\cite{van2014designing}) in isolation. Many of the studies in this research track stay focused on PbD principles (like data minimization~\cite{gurses2011engineering}) without approaching them from an engineering perspective systematically. Some of the studies on PbD highlighted the impact that implementation of PbD should have across different SDLC phases (e.g.,~\cite{kung2011privacy, notario2015pripare}). For example, Notario et al.~\cite{notario2015pripare} offered an approach integrating goal-oriented and risk-based approaches linking requirements analysis, risk management, design, and verification phases. However, this and similar approaches focus on the alignment of different activities across the SDLC and only superficially addresses GDPR as a source of PbD requirements systematically. Some studies that consider GDPR specifically (e.g., ~\cite{mashaly2022privacy}) emphasize the importance of bridging the gap between regulatory requirements and software architecture. However, they have not suggested any systematic way of addressing R\&S specifications in their conjunction. We consider some of such studies in the following subsection.

The recent studies focusing on PbD and/or GDPR compliance implementation suggested technical approaches (especially in the domains of Internet of Things and cloud-based solutions)~\cite{barati2020tracking,cambronero2024towards}, or developed ontologies supporting PbD (e.g.,~\cite{alkhariji2023semantics,zhou2023compliance}). However, such studies covered only particular parts of GDPR (e.g., accountability principle~\cite{zhou2023compliance}) or have not made an explicit link to GDPR norms covered by such approaches (e.g.,~\cite{alkhariji2023semantics}). Only rare studies detailed how they developed a suggested solution, and among those, only some suggested a structured approach. For example, Zhou et al.~\cite{zhou2023compliance} provided a requirements analysis result and a domain model; however, the applicability of the method to other GDPR norms and the applicability of the method by anyone other than the authors was not explored.

\paragraph{Requirements and System Specification for GDPR Compliance}\label{sec:specificationCompliance}
Only some studies considered both R\&S specifications for PbD. The majority of these few studies again mainly focus on the RE or SDA perspective, and do not address the relationships between them specifically. For example, Tamburri ~\cite{tamburri2020design} has suggested a formal concept analysis method for requirements specification and has derived three general design principles for GDPR compliance which are only marginally relevant inputs for the SDA phase. Ayala-Rivera et al.~\cite{ayala2024gdpr} offered a technique to implement GDPR compliance controls, however, have not suggested a method for requirements specification. Some studies have reused the results of previous studies to produce requirements or system specifications. For example, Andrade et al.~\cite{andrade2024privacy} mapped PbD principles to privacy patterns previously cataloged in another study. Such an approach has not made the connection between requirements and system specifications fully explicit. In many cases, the derivation of requirements and system specifications was not transparent enough because corresponding methods have not applied any systematic approach to processing GDPR.

Some SE studies suggest approaches to the relationship between R\&S specifications outside the context of PbD (e.g., Twin Peaks model~\cite{nuseibeh2001weaving}). However, none of these studies is useful or was systematically adapted for PbD.

\paragraph{Modeling Regulations for GDPR Compliance}\label{sec:regulationModeling}
Many studies focus on modeling GDPR for legal informatics and RE purposes without applying such models for system specification purposes(e.g.,~\cite{torre2021modeling}). Other studies focus on R\&S specification without addressing the relationship and traceability to the original GDPR text (e.g.,~\cite{lamari2022process}). AI\&Law research (often applicable for RE purposes too) is mainly focused on modeling regulations (usually in the form of ontologies) for the purposes of information retrieval, automated reasoning, or compliance checking (e.g., ~\cite{robaldo2024compliance, palmirani2018pronto}). Not all of the studies explicate the method for processing the GDPR text systematically to derive corresponding models. This does not allow to address the demands for reconciling legal and engineering perspectives and reproducing model synthesis to address regulatory changes or processing of multiple regulations simultaneously.

\paragraph{Summary}
There is a clear knowledge gap on the relationship between R\&S specifications in the implementation of PbD and the ways how GDPR can be systematically approached to derive R\&S specifications.

\begin{landscape}
\begin{figure}
    \centering
    \includegraphics[width=1\linewidth]{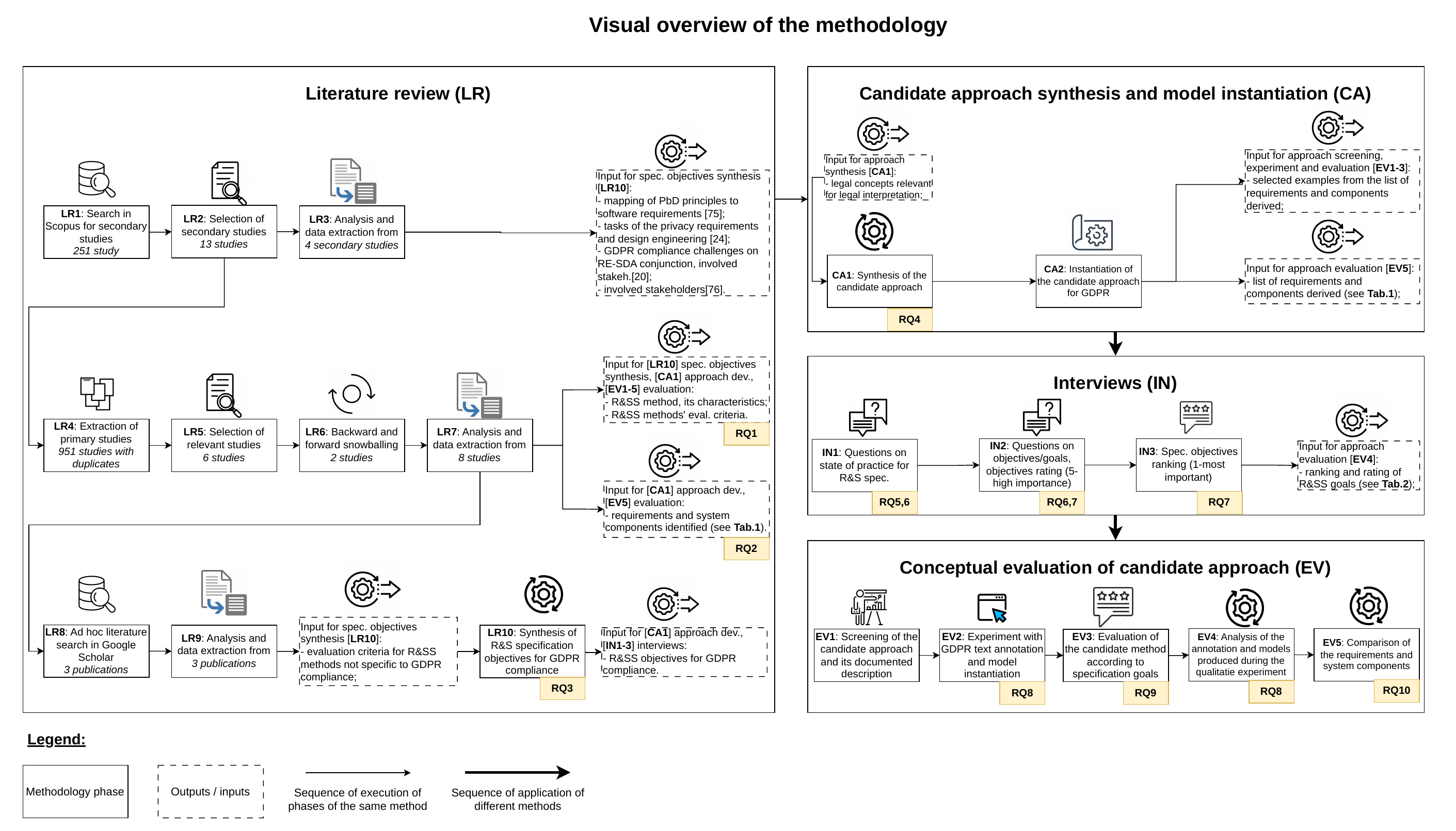}
    \caption{Summary of the methodology applied in this study. We annotate each step of the method execution and refer to them in the text of the manuscript.}
    \label{fig:methodology}
\end{figure}
\end{landscape}

\section{Research Methodology}\label{sec:methodology}
In this section, we describe the methods we applied to execute our study (see Figure~\ref{fig:methodology} for a visual overview). First, we conducted a literature review (LR) to identify the R\&S specification objectives and synthesized our candidate approach (CA). Next, we conducted semi-structured interviews (IN) with the roles involved in R\&S specification for GDPR compliance using the R\&S specification objectives. Finally, we conducted an evaluation of our candidate approach (EV) using the R\&S specification objectives.

\subsection{Literature Review}\label{sec:methodologyLr}
To execute the literature review, we followed the guidelines of Kitchenham~\cite{kitchenham2007guidelines} and guidelines on snowballing by Wohlin~\cite{wohlin2014guidelines}. We formulated the following research questions for the literature review:

\textbf{RQ1:} What is the state of the research of PbD and R\&S specification for GDPR compliance? \textbf{RQ1.1:} What are the available specification methods? \textbf{RQ1.2:} What are the evaluation criteria of such methods?

\textbf{RQ2:} What are the requirements and system components derived with existing R\&S specification methods?

\textbf{RQ3:} What are the specification objectives that need to be achieved in R\&S specification?

We executed the literature review in three phases. We started with the tertiary review in order to cover multiple existing secondary studies and their analysis results. Additionally, multiple studies beyond the research area of GDPR compliance and PbD considered the R\&S specifications evaluation criteria. We executed the following three reviews, a tertiary study of existing secondary studies (LR1-LR3), a secondary study of studies selected in secondary studies (LR4-LR7), an ad hoc literature review to complement the results of the previous phases (LR8-9). We completed our literature review by synthesizing R\&S specification objectives for PbD (LR10). The detailed description of each step is provided next.

\paragraph{LR1: Search of relevant secondary studies}
First, we executed the following search query in Scopus (without  filters applied) and obtained 251 results:
\begin{quote}
    ``software'' AND (``compliance'' OR ``privacy by design'') AND (``secondary study'' OR ``literature review'' OR ``mapping study'')
\end{quote}

\paragraph{LR2: Selection of relevant secondary studies}
After checking the titles and abstracts we selected 13 secondary studies relevant to the topic of R\&S specification for GDPR compliance or PbD.

\paragraph{LR3: Analysis and data extraction from the selected secondary studies}
The first author of the study read through the secondary studies to identify and extract the data on the topic of R\&S specification for GDPR compliance.

\paragraph{LR4: Extraction of primary studies from the selected secondary studies}
Considering that the review of secondary studies has not provided significant results on R\&S specification methods or their evaluation criteria, we selected and analyzed the primary studies used in the 13 identified secondary studies.

\paragraph{LR5: Selection of primary studies}
We used the following inclusion criteria to select primary studies:
\begin{itemize}[nosep]
    \item[IC1]study considers relationships between RE and SDA in the context of GDPR compliance or PbD; and
    \item[IC2]study reports both R\&S specifications or evaluation criteria for R\&S specification methods for GDPR compliance or PbD; and
    \item[IC3]study is in English; and
    \item[IC4]study is peer-reviewed.
\end{itemize}

\paragraph{LR6: Backward and forward snowballing}
Based on the initial seed of the relevant primary studies, we conducted forward and backward snowballing~\cite{wohlin2014guidelines} using the same selection criteria.

\paragraph{LR7: Analysis and data extraction from primary studies}
Next, we extracted and analyzed from the selected primary studies the data on characteristics of R\&S specification method and its evaluation criteria, and  requirements and system components derived with the method.

\paragraph{LR8: Ad hoc literature review}
As the previous literature review phases did not result in any specification objectives and evaluation criteria for specification methods for PbD, we additionally executed an ad hoc search in Google Scholar using the keywords from the original search query.

\paragraph{LR9-10: Synthesis of specification objectives}
Next, we conducted a thematic analysis of the collected data according to Braun\&Clarke~\cite{braun2006using} and applied meta synthesis~\cite{sandelowski2006handbook} to synthesize the R\&S specification objectives. The first author, after familiarizing with the collected data, generated an initial set of codes to identify relevant data (e.g., specification tasks, challenges). After that, the first author created an initial set of themes (e.g., ``capturing legal knowledge''). During the review of the themes, the first author identified the relationships between different themes, and refined them (e.g.,  merged together related themes). Finally, names, definitions and descriptions of the synthesized R\&S specification objectives were documented.

\subsection{Candidate Approach Synthesis and Model Instantiation}\label{sec:methodologySynthesis}
The synthesis of our candidate approach was driven by the following research question:

\textbf{RQ4:} What are the requirements for an approach to support R\&S specification for PbD?

To answer RQ4, we applied a meta-synthesis approach~\cite{sandelowski2006handbook} on the previously collected data. First, we analyzed synthesized specification objectives, and methods' evaluation criteria to derive the very first list of requirements to R\&S specification approach. Next, we analyzed existing methods, together with requirements and system components derived with these methods. We focused on how existing methods can address specification objectives (e.g., capture legal knowledge). Next, we formulated a final list of requirements (see Section~\ref{sec:resultsSynthesis}). We used previous studies to identify the legal concepts relevant to R\&S specification for PbD (e.g.,~\cite{kosenkov2024developing}).

According to the requirements, we synthesized our candidate approach (CA1 in Figure~\ref{sec:resultsSynthesis}, applied it to the text of GDPR and developed the content model of R\&S specifications (CA2 in~\ref{fig:methodology})).
We obtained a list of requirements and system components (Table~\ref{tab:reqsComps}) that we used later as a ground truth for candidate approach screening, and experiment during the evaluation (EV in~\ref{fig:methodology})). The first author of the study developed and applied the candidate approach, drawing on his legal domain knowledge acquired through a law degree and professional certifications in European data protection law.

\subsection{Semi-Structured Interviews}
\label{sec:methodologyInterviews}

We conducted semi-structured interviews following the guidelines of Linaaker et al.~\cite{linaaker2015guidelines} and Runeson et al.~\cite{runeson2009guidelines} to describe the practice of R\&S specification for PbD. The following research questions guided the interviews process:

\textbf{RQ5:} How do practitioners approach R\&S specification for PbD?

\textbf{RQ6:} What are the specification objectives for implementing PbD?

\textbf{RQ7:} How do practitioners rate and rank the importance of the specification objectives?

\paragraph{Participant Selection} We applied purposive sampling with snowballing to recruit participants. Our call for participation targeted engineering roles (e.g., architects) and roles collaborating with the engineers (e.g., legal experts) during R\&S specification for GDPR compliance. We asked that interviewees had technical insights and experience with GDPR. We conducted interviews with twelve persons representing eight companies of various sizes (see~\ref{tab:participants}). The interviewees agreed to participate under conditions of anonymity, and hence they did not disclose any information about their companies.

\begin{table*}[ht]
\tiny
\renewcommand\theadfont{\tiny}
\setlength{\tabcolsep}{3.75pt}
\centering
\rowcolors{3}{gray!15}{white}
\begin{tabular}{ccccc}

\multicolumn{1}{c}{\thead{\tiny{ID}}} & \multicolumn{1}{c}{\thead{\tiny{Company}}} &  \multicolumn{1}{c}{\thead{\tiny{Role}}} & \thead{\tiny{Exp.}} & \thead{\tiny{GDPR Exp.}} \\ \hline
I1 & C1 & Tech lead & 4 & 3 \\ 
I2 & C1 & Sales engineer & 4 & 6 \\ 
I3 & C1 & Stream lead & 28 & 6 \\ 
I4 & C1 & Data engineer & 3,5 & 3,5 \\ 
I5 & C2 & Web marketing spec. & 8 & 5 \\ 
I6 & C3 & Security manager & 34 & 7 \\ 
I7 & C4 & Data engineer & 4 & 2 \\ 
I8 & C5 & IT Compl.\&Audit Head & 30 & 7 \\ 
I9 & C6 & Cloud Sec. Architect & 14 & 5 \\
I10 & C7 & Architect \& Req-s Eng. & 7 & 6 \\
I11 & C8 & Project manager & 20 & 7 \\
I12 & C8 & Tech. project manager & 3 & 3 \\
\hline
\end{tabular}
\caption{Overview of demographic data (ID, Company, role, general experience, experience with GDPR) of interviewees and approach evaluation participants.}
\label{tab:participants}
\end{table*}

\paragraph{Data Collection} To structure the interview process and data analysis, we adapted the Goal Question Metric (GQM) approach~\cite{caldiera1994goal,van2002goal}. The GQM is a structured approach to defining and measuring software quality across three levels of measurement which are: conceptual level defining goals to be achieved, operational level at which questions are defined for evaluating the goals' achievement, and quantitative level defining metrics or data required to answer the questions. For each R\&S specification objective (denoted as ``classes of goals'' during the interviews), we formulated open-ended questions for each level of the GQM. We first asked interviewees about concrete \emph{goals}---the purposes for which the outcomes of R\&S specifications are used and which drive the specification. Next, we asked interviewees to formulate \emph{questions} that need to be addressed in connection with specification goals. Finally, we asked interviewees about \emph{metrics} that could be applied to measure the specification goal achievement and/or answer the formulated questions.

During the interviews, we used closed-ended questions to rate the importance of each of the five specification objectives on a scale from 1---low importance to 5---high importance. At the end of each interview, we asked interviewees to rank the specification objectives from 1---most important to 5---least important, with each ranking assigned only once. The difference in the scale used for rating (1---low importance) and ranking (1---most important) was introduced intentionally to get more detailed reflections from interviewees on the specification objectives and check for results consistency.

The interviews were conducted by the first author (ten interviews) and the second author (two interviews) of this paper. Two interviewers were involved because of the limited availability of interviewees and the need to conduct interviews simultaneously. All the interviews involved one interviewer and one interviewee and were conducted offline (five interviews) or online(seven interviews). 
In all cases, interviews were recorded using Microsoft Teams or Zoom video conferencing software.

\paragraph{Data Analysis and Synthesis}
The first author transcribed the interviews and used the tool Taguette for coding. We applied deductive coding based on the topics of the interview questions (specification objectives, their importance, specification goals, questions, and metrics). We annotated other relevant data with a separate code, for which, however, no new themes were identified.  
We performed member checking with the interviewees by sending them a summary of the interview results via email.

To identify how R\&S specifications conjointly contribute to the achievement of specification objectives, we looked for direct statements from the interviewees. Next, we analyzed if the specification goals identified by interviewees are related to one of specifications, or their intersection (e.g., the modular design was considered as mainly related to system specification).

Taking into account that only two interviewees (I1;I11) used reported questions and metrics in practice, we limited our analysis to the goals and did not analyze questions and metrics.

\subsection{Candidate Approach Evaluation Methodology}\label{sec:methodologyEvaluation}

For the evaluation of our candidate approach, we formulated the following research questions:

\textbf{RQ8:} To what extent can practitioners apply our proposed approach to create a model of R\&S specifications content for PbD?

\textbf{RQ9:} How do practitioners evaluate our proposed approach?

\textbf{RQ10:} How do the requirements and system components, derived using our candidate approach, differ from ones derived with existing methods?

To evaluate our candidate approach, we followed the guidelines by Kitchenham~\cite{kitchenham1996desmet} and Wohlin~\cite{wohlin2012experimentation}, combining screening (evaluation based on documentation) and experiment for the  analysis of our candidate approach.

We recruited evaluation participants from the interviewees to make sure that we have data about the participants' understanding of the R\&S specification objectives. The evaluation was conducted with nine participants immediately after the interview following three steps.

First, for the screening of the approach, we provided the participants hand-outs with the approach description, examples of articles annotations and the resulting specifications' content model. The first author of the study explained the information and answered questions from participants.

Next, for the experiment, we asked participants to complete two tasks and annotate excerpts from two other GDPR articles, and add new elements to the resulting R\&S specification content model. Completion of the model was voluntary, and eventually, five out of nine participants specified the model. Participants used the hand-outs with the approach description and examples during the experiment. Next, we demonstrated to participants solutions to the tasks---annotations of excerpts and the resulting model developed by the first author of this study in CA2~\ref{fig:methodology}. We asked participants if they had any comments or questions about the solutions. Finally, we asked the participants to evaluate the approach according to the usefulness of achieving the five specification objectives discussed during the interviews on a scale from 5---useful to 1---not useful (RQ9).
To identify the extent to which practitioners can apply our approach (RQ8), we analyzed the annotation and modeling results and compared them to the ground truth developed by the first author of the study (Table~\ref{tab:experiment}). For this, we numbered the annotations (10 annotations) and model elements (6 elements) in the ground truth.

To answer RQ10, we compared the requirements and system components derived using our candidate approach with requirements and system components that were reported in the studies identified in the literature review.

\section{Results}\label{sec:results}
In this section, we present the results of the literature review (LR), candidate approach synthesis (CA), interviews (IN) with a focus on the five R\&S specification objectives, and candidate approach evaluation (EV).

\subsection{Literature Review Results}\label{sec:resultsLr}
Here, we report the results of the three phases of the literature review and the synthesized specification objectives.

\paragraph{Tertiary Study on R\&S Specification \& PbD (LR1-3)}
Existing secondary studies have not specifically focused on R\&S specification for GDPR compliance or RE and SDA relationship, but rather highlighted it in the context of their research scope. Among the 13 secondary studies we selected, the majority of the studies were focused on PbD (e.g.,~\cite{caiza2019reusable}), compliance of software systems to different regulations including GDPR (e.g.,~\cite{mubarkoot2023software}, specifically focused on GDPR or privacy in RE (e.g.,~\cite{herwanto2024towards}). Many secondary studies pointed out that primary studies considered both RE and SDA phases of SDLC in the context of GDPR compliance and/or PbD (e.g.,~\cite{kosenkov2024systematic}).

We identified and extracted the following relevant data from four secondary studies: tasks of the privacy requirements and design engineering from~\cite{herwanto2024towards}, mapping of PbD principles to software system requirements from~\cite{saltarella2024translating}, challenges to compliance and concerned stakeholders from~\cite{mubarkoot2023software} and from~\cite{kosenkov2024systematic}.
 
\begin{colora}
\small
\textbf{RQ1:}
\textit{Secondary studies highlight the importance of R\&S specification or the RE and SDA relationship in the context of PbD and GDPR compliance. However, none of the studies provides an in-depth analysis of these topics.
}
\end{colora}

\paragraph{Secondary Study on R\&S Specification for GDPR Compliance (LR4-7)}
We report the results of the review of the primary studies selected from the data of the secondary studies considered in the previous literature review phase.

\paragraph{R\&S Specification Methods}
We identified five primary studies that considered R\&S specification and reported both R\&S specifications~\cite{hjerppe2019general, shah2019analyzing, tomashchuk2020operationalization, istvan2021software, stefanova2021privacy} (see Table~\ref{tab:reqsComps} for an overview of requirements and system components). Many primary studies were not selected because they reported only requirements or only system components (e.g.,~\cite{ayala2024gdpr}).

None of the studies reported a systematic R\&S specification method and the reported requirements-to-system components mappings were ad hoc. Hjerppe et al.~\cite{hjerppe2019general} reused an existing method (twin peaks model) in combination with additional sources of knowledge and interaction with experts. Tomashchuk et al.~\cite{tomashchuk2020operationalization} primarily relied on an interaction with legal experts and emphasized the importance of considering tradeoffs during such interaction. The remaining studies followed a completely ad hoc approach. None of these studies suggested evaluation criteria for the applied methods.

All studies have used natural language to specify requirements. However, approach to system specification was different and included multiple approaches such as design choices~\cite{hjerppe2019general}, features~\cite{shah2019analyzing}, functionality and identified impact on existing system components~\cite{istvan2021software}, architectural approach~\cite{stefanova2021privacy} and simplified reference architecture~\cite{tomashchuk2020operationalization}.

\begin{colora}
\small
\textbf{RQ1.1:}
\textit{No primary study used a systematic, but rather an ad hoc method for R\&S specification for GDPR compliance or mapping of requirements and system components.}
\end{colora}

\paragraph{R\&S Specifications Derived}
Selected primary studies derived a different number of requirements (from 5 to 13) and system components (from 5 to 10). Overall, R\&S specifications show some incoherence across methods potentially related to the absence of systematicity in approaches (e.g., demands related to data subject consent are addressed with requirements specification (R4) in~\cite{istvan2021software, shah2019analyzing}, with system specification (C3) in~\cite{hjerppe2019general}, and with both requirements (R3, R9) and system specification (C1) in~\cite{tomashchuk2020operationalization}). Only one study mapped the relationships between the identified requirements~\cite{hjerppe2019general}. Three studies~\cite{istvan2021software, tomashchuk2020operationalization, stefanova2021privacy} mapped the relationships between the derived system components. Finally, three out of five studies provided a mapping between GDPR articles and requirements~\cite{hjerppe2019general, shah2019analyzing, istvan2021software}.

\begin{table*}
\centering
\resizebox{1\textwidth}{!}{%
\footnotesize
\centering
	\begin{tabular}{p{1.25cm}p{5.75cm}p{4.55cm}p{4.5cm}p{5.5cm}p{4.25cm}} \toprule  
		\thead{} & \thead{Candidate approach in this study} & \thead{Hjerppe~\cite{hjerppe2019general}} & \thead{Tomashchuk~\cite{tomashchuk2020operationalization}} & \thead{Istvan~\cite{istvan2021software}/Shah~\cite{shah2019analyzing}} & \thead{Stefanova~\cite{stefanova2021privacy}}
            \\\midrule
            Requir. & \begin{enumerate}[label=\bfseries R\arabic*:,leftmargin=*,labelindent=0em,labelsep=*, align=left,nosep]
\item processing lawfullness;
\item data subject's consent;
\item processing fairness;
\item processing transparency;
\item specified and explicit purpose;
\item purpose-specific processing;
\item adequacy of processing purpose;
\item data accurate and up-to-date;
\item limited data subject identification;
\item concent clarity and demonstrability;
\item consent explicitness;
\item timely information provision;
\item timely data erasure;
\item appropriate measures;
\item security of data processing. \end{enumerate} & 
\begin{enumerate}[label=\bfseries R\arabic*:,leftmargin=*,labelindent=0em,labelsep=*, align=left,nosep]
\item isolation of different processes of individual purposes;
\item encryption in data storage and transport; \item state-of-the-art methods, libraries, and protocols;
\item pseudonymization while it is possible;
\item in-process separation of data from identifiable people;
\item detailed risk documentation and avoidance;
\item access to machine-readable data set for a single person;
\item documentation of the process in full stack.
\end{enumerate}
&
\begin{enumerate}[label=\bfseries R\arabic*:,leftmargin=*,labelindent=0em,labelsep=*, align=left,nosep]
\item informing users about processing, controller and processors;
\item informing about processing basis;
\item consent collection;
\item data management and data storage;
\item data correction and deletion;
\item restriction of processing;
\item logging;
\item data export;
\item consent withdrawal;
\item authorization;
\item data integrity;
\item encryption;
\item data loss prevention.
\end{enumerate}
&
\begin{enumerate}[label=\bfseries R\arabic*:,leftmargin=*,labelindent=0em,labelsep=*, align=left,nosep]
\item data collected and used for specific purposes
\item data should not be stored beyond its purpose
\item controller must be able to demonstrate compliance
\item get users’ consent
\item provide users a timely access to data
\item find and delete groups of data
\item transfer data to other controllers upon request
\item data should not be used for any objected reasons
\item safeguard and restrict access to data
\item store audit logs of all operations
\item implement appropriate data security measures
\item share insights and audit trails
\item control where the data resides
\end{enumerate}
&
\begin{enumerate}[label=\bfseries R\arabic*:,leftmargin=*,labelindent=0em,labelsep=*, align=left,nosep]
\item giving consent;
\item consent withdrawal;
\item data access;
\item data correction and removal;
\item data retrieval.
\end{enumerate}
            \\\midrule
System Comp.
&
\begin{enumerate}[label=\bfseries C\arabic*:,leftmargin=*,labelindent=0em,labelsep=*, align=left,nosep]
\item processing purpose management;
\item consent management service;
\item data subject identification;
\item data processing information provision;
\item processing restriction/objection;
\item data erasure or rectification;
\item data correction;
\item data access and copy;
\item data portability;
\item processing and breach notification;
\item pseudonomization;
\item data minimization;
\item recording of processing activities.
\end{enumerate}
&
\begin{enumerate}[label=\bfseries C\arabic*:,leftmargin=*,labelindent=0em,labelsep=*, align=left,nosep]
\item static analysis
\item pseudonimization;
\item consent management;
\item personal data event log;
\item GDPR-request management;
\item core personal data (requests);
\item restriction management.
\end{enumerate}
&
\begin{enumerate}[label=\bfseries C\arabic*:,leftmargin=*,labelindent=0em,labelsep=*, align=left,nosep]
\item Consent center
\item Access manag. / disclosure, logging center
\item Permissions / restrictions center
\item Data export center
\item Read/write service
\item Data correction service
\item Data deletion service
\item Keep data alive service (timely deletion)
\item Database
\item Backup.
\end{enumerate}
&
\begin{enumerate}[label=\bfseries C\arabic*:,leftmargin=*,labelindent=0em,labelsep=*, align=left,nosep]
\item timely deletion
\item monitoring and logging
\item metadata indexing
\item access control
\item encryption
\item data location management
\end{enumerate}
&
\begin{enumerate}[label=\bfseries C\arabic*:,leftmargin=*,labelindent=0em,labelsep=*, align=left,nosep]
\item Data Governance Service (DGS);
\item User Personal Data Access Service;
\item Client Services;
\item Database Service;
\item Authentication Service.
\end{enumerate}
    \\\bottomrule
	\end{tabular}}
 \caption{Overview of requirements and system components derived using our candidate approach and four candidate approaches identified in the literature review}\label{tab:reqsComps}
\end{table*}

\begin{colora}
\small
\textbf{RQ2:}
\textit{Only five studies reported requirements and system components derived using their suggested method. These studies do not discuss in-depth the relationship between R\&S specifications for GDPR compliance, and only some identify the relationships between the system components for compliance.}
\end{colora}

\paragraph{Specification Method Evaluation Criteria}
We identified three studies that have not reported both R\&S specifications, but considered method evaluation criteria in the context of GDPR compliance and PbD~\cite{dewitte2019comparison, sion2020dpmf, sion2019architectural}.

These studies suggested the following criteria that we extracted: support of both legal and architectural modeling paradigms and concepts~\cite{dewitte2019comparison, sion2020dpmf, sion2019architectural}; the construction of legal argumentation~\cite{sion2020dpmf} and availability of information for legal activities~\cite{sion2019architectural}; verification and validation of model soundness and legal assessments (data protection impact assessment)~\cite{sion2020dpmf}; generation of appropriate accountability documentation~\cite{sion2020dpmf}; model management over time~\cite{sion2020dpmf};  support soundness criteria~\cite{sion2019architectural}; evaluation of GDPR requirements~\cite{sion2019architectural};  support for risk management~\cite{sion2019architectural}.

\paragraph{Ad hoc Literature Review Result (LR8-9)}\label{sec:adhocLR}
As we have identified only three studies considering evaluation criteria for R\&S specification methods, we conducted an additional ad hoc literature search to explore further approaches for this purpose. We executed a search in Google Scholar using keywords used in the previous phases of the review which resulted in three publications.

Marques\&Yelisetty~\cite{marques2019analysis} suggested seven main attributes for R\&S specification in regulated environments. The technical report by Bass et al.~\cite{bass2006comparison} suggested nine criteria for comparing requirements specification methods from a software architecture perspective and has not considered regulatory compliance in this context. Galster et al.~\cite{galster2009comparing} suggested 31 criteria for comparing methodologies for the transition between software requirements and architectures not specific to regulatory compliance. The most mentioned evaluation criteria for R\&S specification methods in these publications were traceability, consistency, identification of architecture-relevant requirements, facilitation of activities, and stakeholder participation.

\begin{colora}
\small
\textbf{RQ1.2:}
\textit{Studies on GDPR or PbD mainly emphasize two criteria for the evaluation of R\&S specification methods for GDPR compliance: support for architectural and legal domain concepts, and support of legal activities and documentation. Studies not focusing on GDPR compliance also suggest some useful general evaluation criteria such as traceability, consistency.}
\end{colora}

\paragraph{Synthesis of R\&S Specification Objectives for GDPR Compliance (LR10)}\label{sec:synthesisLR}
Next, we used the extracted data to synthesize five R\&S specification objectives (SO) for GDPR compliance, which we report next.

\textbf{SO1: Capturing legal domain knowledge and goals} is required for the effective incorporation of GDPR demands into R\&S specification. Specification methods should ensure the effective identification of architecturally significant requirements~\cite{bass2006comparison, galster2009comparing}, reflection of legal concepts~\cite{dewitte2019comparison, sion2020dpmf} and definition of privacy goals~\cite{herwanto2024towards}. A better reflection of legal domain knowledge in specifications can be seen as important to overcome the challenges of GDPR abstractness~\cite{kosenkov2024systematic} and the gap between engineering and legal perspectives~\cite{mubarkoot2023software}.

\textbf{SO2: Traceability \& Consistency of Specifications} are traditionally considered to be essential in regulated~\cite{marques2019analysis}  and general SE~\cite{marques2019analysis}. They can be seen as essential to address challenges of certification and compliance provability~\cite{mubarkoot2023software, kosenkov2024systematic}. In the context of PbD, three types of traceability can be identified: (1) from GDPR to requirements specification;(2) from requirements to system specification, and (3) from GDPR to system specification.

\textbf{SO3: Separation of Compliance \& Non-Compliance Concerns} is potentially important in R\&S specification because regulated data or system components can require specific handling~\cite{herwanto2024towards, saltarella2024translating}. Herewith, separation of concerns can be essential to address such challenges as conflicts with existing SE methods~\cite{kosenkov2024systematic} or outsourcing of software development~\cite{mubarkoot2023software}.

\textbf{SO4: System specification transparency \& Overview} as part of the documentation process can support accountability and provability of system compliance~\cite{sion2020dpmf}. Specifications (re)use can facilitate the involvement or informing stakeholders~\cite{galster2009comparing, saltarella2024translating, mubarkoot2023software, kosenkov2024systematic}, and support PbD-related activities of other stakeholders (e.g., risk management~\cite{sion2019architectural, herwanto2024towards}.

\textbf{SO5: Specification enabling system flexibility} can be important to avoid ``hardwiring'' of compliance potentially limiting their usability~\cite{lopez2020business}. System flexibility can be important to rapidly and effectively react to changes both in regulations and software systems~\cite{kosenkov2024systematic} and maintain models developed for PbD implementation over time~\cite{sion2020dpmf, sion2019architectural}.

\begin{colora}
\small
\textbf{RQ3:}
\textit{We synthesized the following five specification objectives: capturing legal domain knowledge and goals, traceability \& consistency of specifications, separation of compliance \& non-compliance concerns, system specification transparency \& overview, specification enabling system flexibility.}
\end{colora}

\subsection{Candidate Approach (CA)}\label{sec:resultsSynthesis}
Next, we describe our candidate approach for integrated R\&S specification for PbD developed on the basis of the five R\&S specification objectives described above and analysis of existing R\&S specification methods.

Our approach is built on the idea that R\&S specifications for the implementation of PbD should be developed on the basis of the structured and precise reflection of GDPR content. GDPR text should serve as a source of the domain knowledge and a reference point for R\&S specifications. Capturing GDPR content across different levels of abstraction should preserve the original intentions of GDPR. Still, decoupling problem and solution demands is required to enable effective SE. Next, we provide an overview of the requirements that drove the development of the candidate approach.

\begin{itemize}[nosep]
    \item[R1] R\&S specification methods should address both problem space and solution space demands contained in GDPR to effectively capture legal knowledge (SO1) (including relationships between system concerns/controls);
    \item[R2] specification methods should facilitate traceability to the legal concepts in the original text of GDPR (SO1, SO2);
    \item[R3] Implementation of PbD requires consistency (absence of conflicts) between GDPR, requirements, and system specifications addressing SO2;
    \item[R4] Specification methods should support systematic specification of GDPR content into R\&S specifications and facilitate GDPR---requirements---system specification traceability (SO2);
    \item[R5] GDPR-based content models of R\&S specifications should enable specifications' transparency to non-technical stakeholders and execution of their tasks (e.g., data protection impact assessment) (SO4),.
\end{itemize}

\begin{colora}
\small
\textbf{RQ4:} \textit{The main requirements to a R\&S specification method for PbD: capacity to address legal knowledge across both problem and solution space-related GDPR demands, capturing legal knowledge on the level of legal concepts, specifications' consistency, supporting traceability between specifications, and transparency to stakeholders and their involvement.}
\end{colora}

Our candidate specification approach was inspired by an artefact-based requirements engineering approach~\cite{mendez2015artefact} focusing on the content of artefacts specified, rather than on the processes applied. We base our candidate approach to R\&S specification for PbD on the development of a content model which abstracts from the modeling concepts and only scopes the type of information needed. Artifact-based approach can allow a more consistent reflection of legal knowledge (SO1), enable better consistency and traceability between specifications (SO2) and allow involvement of the required stakeholders (SO4) by assuring the availability of the required information. 

Our approach consists of two primary components: (1) conceptual model for GDPR content analysis and R\&S specification content model development; (2) two levels of abstraction at which instances of concepts in the content model are allocated (requirements and system specification).

\begin{figure*}[h!]
  \centering
  \begin{tabular}{  c  c  }
    \begin{minipage}{.3\textwidth}
    \includegraphics[width=1.1\linewidth]{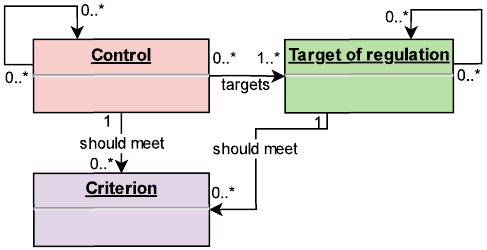}
    \end{minipage}
    &
    \begin{minipage}{.7\textwidth}
    \includegraphics[width=0.9\linewidth]{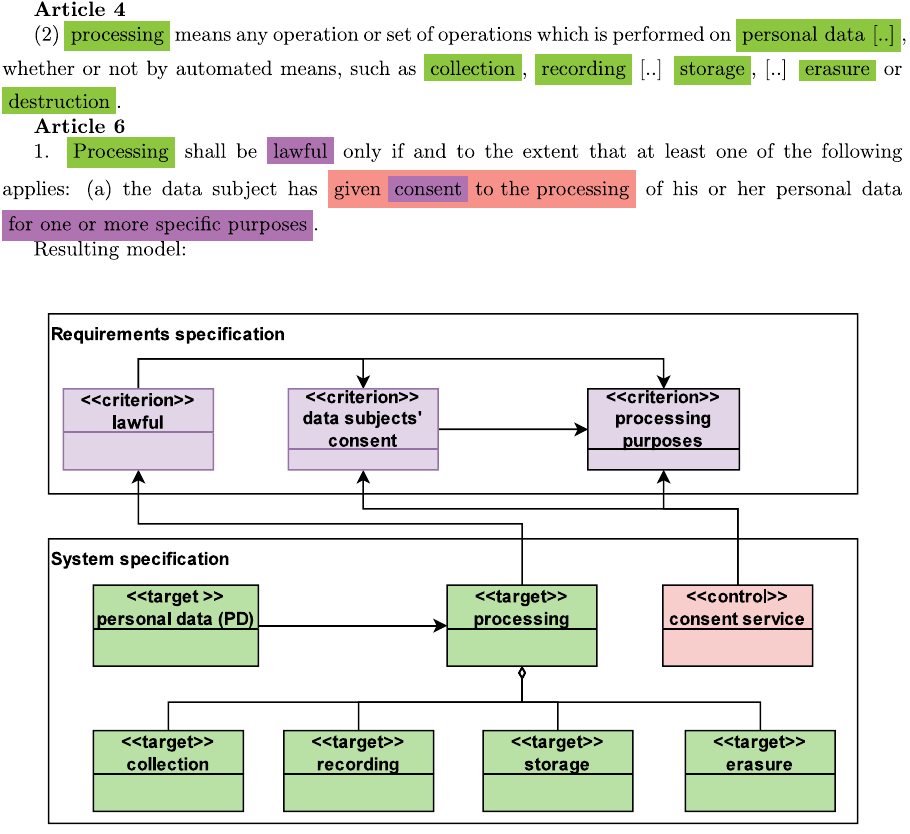}
    \end{minipage}
    \\
  \end{tabular}
  \caption{Left: Conceptual model underlying our approach to R\&S specification for GDPR compliance. Right: Example of the text of excerpts from GDPR Art. 4 and 6 annotated following the conceptual model and the resulting R\&S specification content model.}\label{fig:am4rreInstance}
\end{figure*}

We aim to address the challenge of legal knowledge by using original legal concepts applied by legal experts for the analysis and interpretation of GDPR (see Figure~\ref{fig:am4rreInstance}),  ``legal object'' being central to our model. ``Legal object'' is any tangible or intangible entity involved in a legal relationship or act~\cite{ruiter1997basic, hildebrandt20193}. We intentionally limit our model to this concept because, in our opinion, it is the most relevant for the PbD implementation and for evaluating the conceptual basis of our approach. Other concepts like ``legal subject'' (persons capable of having rights and act~\cite{hildebrandt20193}) have a more complex relationship and indirect impact on the specification for PbD. We plan to cover such concepts and further tool support in future work.

After analyzing requirements and system components in the context of the results of the previous studies (Table~\ref{tab:reqsComps}) and derived the following three concepts for our conceptual model (see Figure~\ref{fig:am4rreInstance}). 
\textit{Target of regulation (subclass of legal object)}---existing components of software systems, organizational processes addressed in regulations. \textit{Compliance control (subclass of legal object)}---new or existing components, processes applied to address the targets of regulation. \textit{Criterion}---properties of compliance controls AND/OR targets of regulations which qualify as acceptable from the legal perspective. This conceptual model forms the core of the structure outline in earlier work~\cite{kosenkov2024developing}, setting it in relation to the larger context of the RE artefacts.

Instances of these concepts can be allocated to two levels of abstraction. \textit{Requirements specification}---includes abstract concepts not specific to system-level specifications and demanding additional interpretation. \textit{System specification}---includes concepts directly related to the system and does not require additional interpretation. Instances of concepts allocated to specifications denote the content of GDPR that should be addressed in each level of abstraction with specific modeling concepts and syntax. In this study, we abstract from the steps of identifying the suitable modeling concepts, considering that they are company and project-specific and require evaluation in real-world scenarios. Figure~\ref{fig:am4rreInstance} shows an example of such an instantiation.

Our approach includes the following steps for the application of the model. First, GDPR text is annotated following the conceptual model. Next, each identified instance is allocated to a particular level of abstraction, and relationships between the instances are inferred from the text of GDPR. This allows for the creation of an initial R\&S specification content model that must be further instantiated for each particular software project. Herewith, instances of concepts allocated to R\&S specifications need to be ``translated'' into corresponding requirements and system modeling concepts to derive project-specific requirements and early software architecture.

The anticipated benefits of our R\&S specification approach are as follows:
\begin{itemize}[nosep]
\item Enabling to capture of the legal goals and knowledge explicitly (SO1);
\item Groundedness in the text of GDPR and traceability to GDPR (SO2);
\item Facilitation of traceability of compliance (SO2);
\item Enabling the specification of both concrete system-specific and abstract requirements-specific statements of GDPR (SO1), thus enabling additional(legal) interpretation when required (SO4);
\item Enabling the involvement of legal experts in the interpretation (SO4).
\end{itemize}

\subsection{Results of the Interviews}\label{sec:resultsInterviews}
Here, we summarize the goals identified by the interviewees (see open data set for a complete list), and the quantitative rating and ranking of the main specification objectives (SOs) (Table~\ref{tab:interviewResults}). We provide goal identifiers [e.g., G1.2] that include the interviewee ID [e.g., 1] and the sequence in which the goal was mentioned [e.g., 2] or interviewee ID only [e.g., I1]. As a starting point during the interviews, all the interviewees rated the role of consistency between R\&S specifications for GDPR compliance as high (4 or 5) and considered it a cornerstone for compliance.

\subsubsection{Existing Tools \& Methods}
Only I8 reported that they used a specific tool for GDPR compliance. Other interviewees were reusing existing standard tools and methods, and approaching compliance ad-hoc. Some of the approaches included interviews with legal experts [I1], using checklists developed by customers and vendors [I7; I9], applying cybersecurity best practices [I6], and relying on requirements specified in contracts [I6].

\subsubsection{Key Features of a Potential R\&S Specification Method}
Despite the majority of the interviewees did not use specific tools for R\&S specification for GDPR compliance, they named a number of features specific to such tools: supporting GDPR compliance controls implementation [I4; I8; I9], separating, tracking of regulated types of data and raising awareness about them [I2; I4; I11], concretization of GDPR norms [I3; I6], facilitation of understandability and explainability of GDPR [I10], support of assuring completeness (e.g., checklist) [I8], establishing interaction and common ground between the stakeholders [I1], guidelines to identify what is regulated and to enable better decision-making [I11].

\begin{colora}
\small
\textbf{RQ5:} \textit{The majority of practitioners do not use any specialized methods for R\&S specification for PbD. Rather, practitioners approach it ad hoc and through the reuse of existing methods. Practitioners do not have a clear and systematic vision of the desirable features of R\&S specification tools and/or methods, but identify multiple requirements specific to regulatory compliance.}
\end{colora}

\subsubsection{Goals in R\&S Specification for GDPR Compliance}
Next, we provide a summary of the goals mentioned for each SO.

\paragraph{SO1: Capturing Legal Domain Knowledge \& Goals} The first discernible group of goals related to SO1 was concerned with facilitating understanding of the legal logic in the software design. As high-level legal goals are ``diffused'' [I10] they need to be mapped to something concrete [G2.1]. Clarity of such mapping is important to avoid GDPR misinterpretation [G12.1]. Herewith, it is important to have references and context for GDPR requirements in the future [G11.2] and assure accessibility [G11.3] and understandability of legal knowledge to engineers  [G11.7] (see next).

\begin{colorb}
\small
\textit{``In every [finished] project, it was a while since you did it and you forget why you did it, how to apply it. Since usually the developers don't have legal knowledge, it's always a huge effort. I think [legal knowledge] should be particularly there but in an easy-to-access general way. [In] the next project, you just look at your template, you fill it in, you [..] again remember the context and figure out what to do in this case.''} (I11)
\end{colorb}

Multiple goals were linked to communication and collaboration, such as bringing the required expertise together [G6.3], finding consensus [G6.2], assuring awareness and responsibility [G9.1], or advising clients [G7.2].

Capturing legal knowledge was seen as essential, but rather related to the responsibility of the legal roles [I3], especially as some interviewees [I4; I6] emphasized that they usually interact with legal experts (see next).

\begin{colorb}
\small
\textit{``We do not need that the developer is a GDPR expert. It is assuring that the team brings all the expertise together.''} [I6]
\end{colorb}

Capturing legal knowledge also contributes to the achievement of engineering goals such as addressing GDPR as early as possible [G8.1], understanding design goals and the required flexibility [G4.1-2], implementing of a system according to requirements [G7.1] (see next).

\begin{colorb}
\small
\textit{``It's very important that the legal knowledge is specified into requirements and then a development team can design the system based on it.''} (I3)
\end{colorb}

Other goals were related to auxiliary activities for PbD such as avoiding risks [G3.1], performing gap analysis [G6.4], balancing needs [G6.1], and assessing and spending the right effort [G11.6].

\paragraph{Insights}
SO of capturing legal knowledge depends on the conjunction between R\&S specifications, primarily because of the need to specify legal knowledge into requirements and then system specifications. The relationship between R\&S specifications contributes to the involvement of roles and stakeholders having different responsibilities, and providing a sufficient understanding of GDPR in the design process. Our finding about the demand for the usability of the captured legal knowledge for engineers is somehow aligned with the previous studies emphasizing the gap between engineering and legal perspectives~\cite{bobkowska2010efficient}.

\paragraph{SO2: Traceability \& Consistency}
Goals supported by this SO were related to guaranteeing the correct implementation of GDPR [G2.4; G5.1; G7.3; G8.4; G11.8-10], assuring testability [G10.6] and provability [I3; I5] of compliance. Another group of goals was related to reacting to changes (tracking changes [G10.4; G12.4], and identifying where changes are required [G10.5]).

Two interviewees considered traceability and consistency as essential for governance of GDPR compliance [G6.6; G9.3] identification of the current compliance status and compliance gaps [G9.4-5]. Non-consistency between R\&S specifications could result in ``overshooting'' or ``undershooting'' of compliance, causing financial damages [I3].

Other goals linked to this SO were supporting version control and ownership [G1.3], implementation and change history [G4.3], assuring that processes support compliance implementation [G10.7], or providing proofs [G6.7], and facilitating auditors to trace the implementation [G12.3].

One important finding is that some interviewees (I1; I5; I10) rated different types of traceability differently, emphasizing the varying needs across the roles involved. I1 and I10 rated the traceability between R\&S specifications higher than the traceability between the GDPR text and system specifications. In their opinion, the latter kind of traceability is more relevant for the involved legal roles, who should be responsible for it. I5, as a web marketing specialist, considered GDPR-to-system specification to be more important because it is related to the final compliance status of the system.

According to I3, the implementation of traceability for GDPR compliance faces the challenge of combining technology and legal knowledge (see next).

\begin{colorb}
\small
\textit{``It's hard to find people who could check that traceability because often the [people having] legal knowledge are not the people who know technology and those who know technology have very hard time reading the original text.''} [I3]
\end{colorb}

\paragraph{Insights}
According to the interviewees, traceability between R\&S specifications is instrumental for implementing and proving compliance, reacting to changes, executing governance, and auxiliary tasks. Our results suggest that technical and non-technical (e.g., legal) roles involved in R\&S specification have different traceability needs which are complementary in achieving PbD.

\paragraph{SO3: Separation of Concerns}
Seven interviewees considered GDPR compliance to be the core business concern not requiring separation, but rather integration, with business concerns [I2-6; I9; I10]. Five other interviewees [I1; I7; I8; I11; I12] were of the opinion that GDPR compliance is closely related to business concerns, but should be isolated for better manageability. Notably, although some interviewees [I10] mentioned that isolation of compliance concerns is not required, these concerns should be easily identifiable. 

The reasons for the separation of compliance concerns were to identify the business value of data [I1], manage compliance to different regulations [I8], in different jurisdictions, and/or business operations [I7], fulfilling certification and validation [G11.14], identification of requirements and system components demanding compliance [G8.5; G7.4] or change [G11.15],  facilitation of audits [G8.7], proving compliance [G11.16], compliance visibility through ``living compliance'' [G11.17], avoiding additional GDPR-related constraints [G7.6], assuring maintainability and scalability (G12.5-6).

I11 highlighted that, in some cases, system specification can result in another set of requirements (see next quotation).

\begin{colorb}
\small
\textit{``We built the system without thinking about GDPR. Then, we created policies, and documented the system architecture. We documented the data parts relevant to GDPR, in which system they are used, and why we have them. Eventually from this process, we got other requirements back.''} [I11]
\end{colorb}

\paragraph{Insights}
It was hard for practitioners to discuss the separation of concerns as one of SO. Some interviewees [e.g., I11] changed their opinion and rating of this SO while answering the interview questions.

The relationship between R\&S specification for the separation of concerns is essential because system specification can result in further additional requirements. According to the interviewees considering that the separation of concerns was not required, it was still important to identify requirements and concerns arising from GDPR.

\paragraph{SO4: Transparency \& Communication Over Specifications}
Many interviewees rated the importance of transparency of specification and communication differently. For this reason, we report both rating values in Table~\ref{tab:interviewResults}, transparency as the first value, and communication as the second one.

Again, some goals in this SO were related to the effectiveness of communication and interaction [G3.4; G5.2; G8.9], common understanding between roles [G7.7-8; G8.13; G9.9; G10.10; G12.7], resolving conflicts and balancing limitations [G8.11-12], optimal effort [G12.8], choosing and discussing different solutions [G11.20] and additional compliance-related costs [I6], faster execution of business processes [G5.3]. According to I8, communication is essential for implementing GDPR compliance (see next).

\begin{colorb}
\small
\textit{``When we have problems with GDPR, system design, it's all about miscommunication. It's always different expectations not communicated.''} (I8)
\end{colorb}

I3 described transparency and communication as crucial for proving compliance (see next).

\begin{colorb}
\small
\textit{``[S]ometimes good communication can make you compliant. If you are almost compliant, and you communicate about it well, you can say this is why we make this decision. [I]t is about taking the appropriate risk for the sensitive data. [I]f you communicate about it really well, you prove that you understand the point.''} (I3)
\end{colorb}

Furthermore, goals were related to facilitating activities connected to GDPR compliance such as audits [G3.3; G8.8; G11.22], privacy impact assessment [G4.5], risk management [G8.10; G9.8], compliance testing [G10.9], updates implementation [G11.21], and prevention of data breaches [G12.11]. I1, I4, I6, and I8 mentioned that, despite communication being important, other stakeholders (e.g., auditors) usually need to understand system specifications on a higher level of detail.

According to I6, bringing different abstraction levels together in a system with both technical and organizational controls (denoted by the interviewee as ``socio-technical system'') would be helpful but is challenging in practice.

\paragraph{Insights}
The goals identified in relation to this SO need to be achieved both in R\&S specifications. Our results suggest that roles and stakeholders require different granularity of the information contained in specifications and in particular in specifications other roles are responsible for. It is noteworthy that interviewees in roles responsible for hands-on implementation of GDPR (e.g., tech lead I1) rated the importance of transparency higher than the importance of communication, while interviewees in managerial roles (e.g., security manager I6) rated the importance of communication higher.

\paragraph{SO5: Specification for System Flexibility}
Some of the goals related to this SO were related to directly supporting engineering tasks such as effective implementation of changes [G4.6; G5.4], ease of identifying system components requiring flexibility, changes or updates [G1.10], minimizing costs of changes and updates [G11.23], good architecture of system [G11.24], adaptability and modularity of design [G5.5; G8.14; G10.13; G10.15] or engineering process [G8.15], and identifying unnecessary work [G11.26]. Some goals were related to the reaction to changes, such as monitoring compliance gaps [G9.11], identification of the relevance of new vulnerabilities [G6.10], traceability to support evolution [G10.14], and facilitating recertification [G11.27].

\paragraph{Insights}
No interviewee directly mentioned the relationship between R\&S specifications in the context of this SO. Our further analysis of specification goals shows that this SO is mainly implemented with system specification, however, multiple goals were related to requirements specification or both R\&S specifications (e.g., monitoring compliance gaps [G9.11]).

\paragraph{Other Specification Objectives \& Goals}
Here, we provide an overview of the goals that interviewees mentioned additionally at the end of the interviews. Many of the goals were related to GDPR obligations implementation such as data type identification [G2.7], interpretation correctness [G6.13] assuring personal data protection [G5.7] and data subject rights [G12.14-15], manageability of GDPR duties [G8.17] and non-compliance cases [G8.16], automation of compliance controls [G1.12], preventing personal data breaches [G6.11] and achieving compliance efficiency [G9.12]. Two interviewees mentioned goals related to assuring business needs [G5.6; G7.9]. Other goals included assuring communication [G1.13], synthesis of best practices for compliance [G11.28], and controlling IT infrastructure and its evolution [G6.12].

\paragraph{Insights}
The majority of the goals not related to the five core SOs were connected with the achievement of business or security goals, correctness and/or efficiency of compliance, manageability, and governance of compliance. Our results suggest that such cross-functional concerns as compliance governance and efficiency also demand conjoint R\&S specification. 

\begin{colora}
\textbf{RQ6:}
\textit{According to practitioners, four out of five SOs (SO1, SO2, SO3, and SO4) are achieved to a greater or lesser extent with R\&S specifications conjointly. Traceability is an essential SO for assuring PbD. However, it needs to be supported by (1) specification transparency to address the challenge of the granularity of information required by different stakeholders and (2) reflection of legal domain knowledge to support the understanding of source and target specifications.}
\end{colora}

\subsubsection{Rating and Ranking of Specification Objectives (IN2, IN3)}
We asked interviewees to rate the importance of SOs during the interviews and rank them after the interviews to better understand the priorities of practitioners. We used a different scale for the rating (5---high importance) and ranking (1---most important) to ensure practitioners analyze the tasks.

\begin{table*}[ht]
\tiny
\setlength{\tabcolsep}{4pt}
\centering
\rowcolors{3}{gray!15}{white}
\begin{tabular}{cclcc|ccccc|ccccc|ccccc}

\multirow{2}{*}{\rotatebox{90}{ID}} & \multirow{2}{*}{\rotatebox{90}{Company}} & \multirow{2}{*}{\rotatebox{90}{Role}} & \multirow{2}{*}{\rotatebox{90}{Work Exp.}} & \multirow{2}{*}{\rotatebox{90}{GDPR}} & \multicolumn{5}{c|}{\thead{\tiny{Rating}}} & \multicolumn{5}{c|}{\thead{\tiny{Ranking}}} & \multicolumn{5}{c}{\thead{\tiny{Evaluation}}} \\ 
&  &  &  &  & \thead{\rotatebox{90}{\tiny{SO1}}} & \thead{\rotatebox{90}{\tiny{SO2}}} & \thead{\rotatebox{90}{\tiny{SO3}}} & \thead{\rotatebox{90}{\tiny{SO4}}} & \thead{\rotatebox{90}{\tiny{SO5}}} & \thead{\rotatebox{90}{\tiny{SO1}}} & \thead{\rotatebox{90}{\tiny{SO2}}} & \thead{\rotatebox{90}{\tiny{SO3}}} & \thead{\rotatebox{90}{\tiny{SO4}}} & \thead{\rotatebox{90}{\tiny{SO5}}} & \thead{\rotatebox{90}{\tiny{SO1}}} & \thead{\rotatebox{90}{\tiny{SO2}}} & \thead{\rotatebox{90}{\tiny{SO3}}} & \thead{\rotatebox{90}{\tiny{SO4}}} & \thead{\rotatebox{90}{\tiny{SO5}}} \\ \hline
I1 & C1 & Tech lead & 4 & 3 & 3 & 4(2) & 4 & 4(3) & 4 & 5 & 4 & 2 & 3 & 1 & 5 & 3 & 4 & 2 & 1 \\ 
I2 & C1 & Sales engineer & 4 & 6 & 5 & 5 & 1 & 5 & 5 & 4 & 1 & 5 & 2 & 3 & 3 & 4 & 1 & 5 & 2 \\ 
I3 & C1 & Stream lead & 28 & 6 & 4 & 3.5 & 1 & 4(5) & 2 & 3 & 4 & 1 & 2 & 5 & 4 & 4 & 3 & 4 & 2 \\ 
I4 & C1 & Data engineer & 3.5 & 3.5 & 5 & 5 & 1 & 5(1) & 2.5 & 1 & 2 & 5 & 3 & 4 & 5 & 2 & 1 & 5 & 3 \\ 
I5 & C2 & Web marketing spec. & 8 & 5 & 5 & 2.5(5) & 1 & 5 & 5 & 1 & 2 & 5 & 4 & 3 & 5 & 4 & 2 & 5 & 4 \\ 
I6 & C3 & Security manager & 34 & 7 & 5 & 4 & 1 & 3(4) & 5 & 1 & 2 & 5 & 4 & 3 & - & - & - & - & - \\ 
I7 & C4 & Data engineer & 4 & 2 & 5 & 4 & 5 & 5(3) & 5 & 2 & 1 & 5 & 4 & 3 & - & - & - & - & - \\ 
I8 & C5 & IT Compl.\&Audit & 30 & 7 & 5 & 4 & 5 & 3.5(5) & 2 & 2 & 3 & 1 & 5 & 4 & - & - & - & - & - \\ 
I9 & C6 & Cloud Sec. Architect & 14 & 5 & 5 & 5 & 1 & 5(4) & 3 & 1 & 3 & 5 & 2 & 4 & 5 & 5 & 2 & 4 & 4 \\
I10 & C7 & Architect\&Req-s Eng. & 7 & 6 & 5 & 5(2) & 1 & 4(5) & 3 & 1 & 2 & 5 & 3 & 4 & 2 & 5 & 1 & 4 & 5 \\
I11 & C8 & Proj. manager & 20 & 7 & 4 & 5 & 4 & 3(5) & 4 & 1 & 2 & 3 & 4 & 5 & 5 & 5 & 4 & 5 & 5 \\
I12 & C8 & Tech. proj. manager & 3 & 3 & 4 & 4 & 4 & 5(4) & 4.5 & 1 & 3 & 4 & 2 & 5 & 5 & 5 & 4 & 5 & 5 \\
\hline
\hline
\multicolumn{5}{l|}{\textbf{Median}} & \textbf{5} & \textbf{4(4)} & \textbf{1} & \textbf{4.5(4.5)} & \textbf{4} & \textbf{1} & \textbf{2} & \textbf{5} & \textbf{3} & \textbf{4} & \textbf{5} & \textbf{4} & \textbf{2} & \textbf{5} & \textbf{4} \\
\hline
\multicolumn{5}{l|}{Mode} & 5 & 4,5(5) & 1 & 5(5) & 5 & 1 & 2 & 5 & 2,4 & 3,4 & 5 & 5 & 4,1 & 5 & 5 \\
\hline
\end{tabular}
\caption{Overview of demographic data (ID, Company, role, general experience (years), experience with GDPR (years)), results of the interviews (Ranking, Importance) and candidate approach evaluation (Evaluation).}
\label{tab:interviewResults}
\end{table*}

\begin{colora}
\small
\textbf{RQ7:}
\textit{The importance SOs is as follows (from the most to the least important with median ranking / median rating): SO1: capturing legal knowledge (1/5), SO2: traceability and consistency (2/4), (3) SO4: specification transparency (3/4.5), (4) SO5: facilitation of system flexibility (4/4), (5) SO3: separation of compliance and non-compliance concerns (5/1).}
\end{colora}

We suggest that SO1 capturing legal domain knowledge received the highest rating and ranking  mainly due to the complexity of legal knowledge. It is noteworthy that SO4 (specification transparency) was rated and ranked practically on par with SO2 (traceability and consistency). SO5 (facilitating system flexibility) and SO3 (separation of concerns) received low median rating and ranking. However, we observed that it was particularly challenging for practitioners to reason about the importance of separation of concerns. 

\subsection{Candidate Approach Evaluation (EV) Results}\label{sec:resultsEvaluation}
In this section, we report (1) the results of the experiment on annotating the text of GDPR and modeling the content of R\&S specification (EV2, EV4) conducted to identify the degree to which our method can be used by practitioners, (2) results of the evaluation approach according to specification goals (EV3) to evaluate the potential usefulness of our method, and (3) comparison of requirements and system components derived with our candidate approach and other methods reported in the literature (EV5) to compare our candidate approach to the ones reported in the literature.

\paragraph{Results of the experiment (EV2, EV4)}
To analyze the results of the experiment we compared the annotations and models produced by participants with the ground truth produced by the first author of the study (see Table~\ref{tab:experiment}).

\begin{table*}[ht]
\tiny
\renewcommand\theadfont{\tiny}
\setlength{\tabcolsep}{3.75pt}
\centering
\rowcolors{3}{gray!15}{white}
\begin{tabular}{c|ccccccccccc|ccccccc}

& \multicolumn{11}{c|}{\thead{Annotations identified}} & \multicolumn{7}{c}{\thead{Components modeled}}\\ 
\thead{ID} & \thead{13.1} &  \thead{13.2} & \thead{13.3} & \thead{13.4} & \thead{13.5} & \thead{15.1} & \thead{15.2} & \thead{15.3} & \thead{15.4} & \thead{15.5} & \thead{A+} & \thead{C1} & \thead{C2} & \thead{C3} & \thead{C4} & \thead{C5} & \thead{C6} & \thead{C+} \\ \hline
I1 & 1 & 1 & 0.7 & 0.8 & 0.7 & 0.9 & 1 & 0.7 & 0 & 0.8 & 3 & - & - & - & - & - & - & - \\ 
I2 & 1 & 0 & 0.9 & 0.7 & 0.7 & 0.7 & 0 & 0.8 & 0 & 0 & 2 & - & - & - & - & - & - & - \\ 
I3 & 1 & 0 & 0.7 & 0 & 0.7 & 0.9 & 0 & 0.7 & 0 & 0.8 & 7 & - & - & - & - & - & - & - \\ 
I4 & 0.8 & 0 & 0 & 0.9 & 0.7 & 0.7 & 1 & 0.7 & 0 & 0 & 2 & - & - & - & - & - & - & - \\ 
I5 & 1 & 1 & 0.9 & 0.7 & 0.7 & 0.9 & 0 & 0.7 & 0 & 1 & 4 & + & - & + & + & - & - & 4 \\
I9 & 0 & 0 & 0.9 & 0.9 & 0 & 0.9 & 0 & 0 & 0 & 1 & 2 & + & - & + & - & - & - & 2 \\
I10 & 1 & 0 & 0.7 & 0.9 & 0 & 0.7 & 1 & 0.7 & 0 & 1 & 1 & + & + & - & - & + & + & 3 \\
I11 & 1 & 0.8 & 0.9 & 0.8 & 0.7 & 1 & 1 & 0.7 & 0.7 & 1 & 7 & + & + & - & - & - & - & 2 \\
I12 & 0.8 & 1 & 0.7 & 1 & 0 & 0.7 & 0 & 0 & 0 & 0 & 1 & + & - & - & - & + & - & 1 \\ \hline \hline
\textbf{Median} & 1 & 0 & 0.7 & 0.8 & 0.7 & 0.9 & 0 & 0.7 & 0 & 0.8 &  &  &  &  &  &  &  &  \\
Mode & 1 & 0 & 0.7,0.9 & 0.9 & 0.7 & 0.9,0.7 & 0 & 0.7 & 0 & 1 &  &  &  &  &  &  &  &  \\






\hline

\end{tabular}
\caption{Results of experiment (EV2) on annotation excerpts of GDPR Articles 13, 15 and modeling of specifications content.\\\hspace{\textwidth}
\tiny{For annotations: 1---annotations in the ground truth was identified completely correctly, 0.9---annotation text was selected completely correctly, but assigned with wrong concept, 0.8---text of annotations was not identified completely correctly, but concept was correctly assigned, 0.7---annotation text was not identified completely correctly and wrong concept was assigned, 0---annotation was not identified. A+---number of extra annotations identified. For system components: system component was identified (+) or was not (-), C+---number of extra system components identified.}}
\label{tab:experiment}
\end{table*}

Out of 90 annotations(9 participants x 10 annotations in the ground truth), 29 were not identified by the participants. Among the 61 identified annotations, 19 were identified completely correctly (1 point), 11 were identified correctly, but assigned with a wrong concept (0.9 points), in 8 annotations text was selected partially correctly, but was assigned with the correct concept (0.8 points), and in 23 annotations text was selected partially correctly and incorrect concepts were assigned (0.7 points). Each participant annotated completely correctly between 1 and 4 concepts, while the rest were identified with at least one mistake in comparison to the ground truth.

On the basis of the analysis of the experiment result, we suppose the following challenges to annotating the text of GDPR. It was hard to address synonyms, which are abundant in GDPR. For example, annotation A13.5 ``the purposes of the processing for which the personal data are intended'' and A15.5 ``the purposes of the processing'' are related to the same <<criterion>> of processing purposes. However, A13.5 was annotated by all participants incompletely and with a wrong concept.

Another challenge was the identification of related concepts. For example, the annotation of the excerpt from Article 15 contains three interrelated instances of concept <<control>>: A15.1 ``obtain from the controller confirmation'', A15.3 ``access to the personal data'' and A15.4 ``[access to] the following information''. All experiment participants identified A15.1, 7 out of 9 identified A15.3, and only one experiment participant annotated A15.4.

Next, we analyzed the model components that five experiment participants developed. Experiment participants correctly specified between two and four of the six components. All participants successfully identified an instance of <<control>> C1, and two participants (I10, I11) identified both instances of <<control>> C1 and C2. However, none of the participants have identified more than two relationships of C1 and C2.

In some cases (I5, I9-I12), we had an opportunity to observe the process of annotation. Overall, it was hard for the experiment participants to annotate text and create models. Participants were often not sure about the correctness of the annotations and models they produced.

\begin{colora}
\small
\textbf{RQ8:}
\textit{Practitioners can use our candidate approach to a limited degree. Our approach provides a basis for the analysis of GDPR and the creation of R\&S specifications' content model, however it is still hard for practitioners to apply, they can miss instances of some of the concepts, identify only part of the required text and/or assign incorrect concepts.}
\end{colora}

\paragraph{Results of the candidate approach evaluation (EV3)}
Next, we provide an analysis of the results of the evaluation of the candidate approach by participants after they screened and applied our candidate approach in the experiment. An overview of the evaluation results is available in Table~\ref{tab:interviewResults}.

The participants positively evaluated the capacity of our approach to achieve SO1 capture legal knowledge (median evaluation of 5---useful). I10 was the only participant assigned an evaluation of 2 (potentially not useful) because, in their opinion, capturing legal domain knowledge is mainly related to ``How to capture legal knowledge?'' and not ``What should be captured?''.

Our candidate approach received a positive median evaluation of 5 (useful) for the achievement of SO4 specification transparency. Here, I10 mentioned that the approach facilitates consistency because concrete concepts were identified directly in the text of GDPR (see next).

\begin{colorb}
\textit{``It gives a good overview of the structures since it's closely related to the original text. If you know how it was derived, it's very transparent. [However] the `Why?' is not covered by your model.''} (I10)
\end{colorb}

In connection with SO2 (traceability), our approach got a median evaluation of 4 (potentially useful). According to I11 our candidate approach creates a basis but does not provide guidance for tracing to the original system specification as the next step. The potential usefulness of our candidate approach in relation to SO5 got a median evaluation of 4 (potentially useful). According to I12 the approach can help to identify what requires changes and enables system flexibility. Our candidate approach received the lowest average evaluation for SO3 (separation of concerns). According to some of the interviewees, our candidate approach allows for the creation of a basis for the separation of concerns, but at the moment does not cover handling of non-compliance (e.g., business) concerns.

\begin{colora}
\small
\textbf{RQ9:}
\textit{Participants evaluated our approach as 5 (useful) in relation to SO1 capturing legal knowledge and SO4 specification transparency, 
as 4 (potentially useful) in relation to SO2 traceability and consistency and SO5 enabling system flexibility. Our approach was evaluated as 2 (potentially not useful) in relation to SO3 separation of concerns.}
\end{colora}

\paragraph{Results of the comparison of the requirements and system components derived (EV5)}\label{sec:resultsComponents}
We compared the requirements and system components derived with our approach to the requirements and system components reported in the studies identified during the literature review (Table~\ref{tab:reqsComps}).
Using our approach, we identified more requirements and system components than with any other previous method (15 requirements in comparison to a maximum of 13 requirements and 13 components in comparison to a maximum of 10 components). One of the benefits of our candidate approach in relation to the requirements derived is that it captures almost all of the core data processing principles formulated in Article 5 of GDPR, while other approaches capture only some of them. However, the requirements captured with our method miss the important accountability principle captured by other methods (R3 in Istvan). Our approach also allows to include relationships between requirements and concrete system components rather than simply listing general requirements (see Figure~\ref{fig:am4rreInstance}). Our approach captures other important relationships as prescribed by GDPR. For example, controls that are prescribed by GDPR are related to particular system components and concerns(e.g., <<control>> ``data access service'' is related to <<target>> ``personal data storage''). Finally, there are potential direct or indirect relationships between different controls (e.g., <<control>> ``information provision service'' needs to access <<control>> ``consent service'' to provide information about purposes for which personal data is processed or access data about such processing purposes generated by <<control>> ``consent service''). In previous studies, relationships between different controls were identified by Tomashchuk~\cite{tomashchuk2020operationalization}, and Stefanova~\cite{stefanova2021privacy}.

Relationships between different system-level components or concerns are hard to capture with requirements specification methods, and hence they need to be captured with (early) system specification. This demonstrates that our approach, focusing on the content of SE artifacts and abstracting from specification methods, is suitable to capture requirements, early system components, and concerns directly from the text of GDPR.

The system components that were identified only with our candidate approach were C1 processing purpose management, C3 data subject identification, C10 processing and breach notification.

\begin{colora}
\small
\textbf{RQ10:}
\textit{Our candidate approach results in the identification of more requirements and system components than with the application of other methods. Our approach allows for establishing relationships between requirements and concrete system components, inferring the relationships between different system components (data processing activities) and required GDPR controls. However, neither our approach nor existing approaches fully duplicate GDPR demands, both in R\&S specifications. This confirms the idea that effective joint consideration of R\&S specifications is required to fulfill GDPR demands.}
\end{colora}

\section{Discussion}\label{sec:discussion}
Privacy by design is a prominent topic in software engineering research. However, only some recent studies address GDPR compliance as a foundation of PbD. Nowadays, the PbD concept is underpinned by the multidisciplinary principles suggested by Cavoukian~\cite{cavoukian2012operationalizing} and Article 25 of GDPR, which formulates PbD from a legal perspective. Herewith, the PbD principle is insufficiently conceptualized from the SE perspective, despite the recent efforts to develop more holistic approaches towards PbD~\cite{herwanto2024towards}. We suggest such an approach by emphasizing the conjoint consideration of requirements and system specifications. Joint consideration of R\&S specifications was recognized as important in some of the recent studies~\cite{herwanto2024towards, kosenkov2024systematic}. However, there are no studies that focused on it or suggested a systematic method for it.

Our interview results reveal that practitioners do not apply any specific methods for R\&S specification for GDPR compliance, with the exception of one interviewee. However, practitioners mentioned some specific needs that may not be fully supported by existing methods. This finding is well-aligned with previous findings indicating that tool support for GDPR compliance is not widely used in practice~\cite{klymenko2022understanding} and that regulatory compliance needs often introduce conflicts with existing SE methods and tools~\cite{kosenkov2024systematic}.

Our interview results confirm that conjoint R\&S specification is required for the achievement of four specification objectives, which are capturing legal domain knowledge, traceability and consistency, separation of concerns, and specification transparency and communication over specification.

Our interview results suggest that for the implementation of PbD, traceability between R\&S specifications needs to be supported with the specification objectives of capturing legal knowledge and specification transparency. This is essential for understanding both source and target specifications and providing the granularity of information required by the stakeholders involved. Previous studies on traceability emphasized the importance of enabling understanding and reasoning about system specifications, reflecting decisions~\cite{cleland2013decision}, and synchronizing different stakeholders~\cite{turban2009bridging}. Our results confirm the essential character of these tasks for PbD implementation.

Our findings are aligned with the idea that regulations are intentionally abstract and need to be interpreted throughout the SDLC, with different specifications and roles involved. This echoes the results of previous studies, pointing to the challenges that different SE roles face with understanding and approaching PbD in the context of their work~\cite{bu2020privacy}.

Our candidate R\&S specification approach, built on the idea of modeling GDPR content using the original legal concepts, was evaluated as useful for reflecting legal knowledge and facilitating specification transparency. Our candidate approach was evaluated as potentially useful for traceability and consistency of specification and system flexibility. The benefits of our approach are mainly related to its capacity to capture the content of GDPR text. Among other benefits, we anticipate that R\&S specifications defined using our approach could be directly mapped to specific norms of GDPR. This, in turn, would allow the identification of requirements or system components with the highest importance based on the number of GDPR norms they map to. Herewith, identification of the system components impacted by the GDPR norms could support the review of existing design practices and facilitate risk assessment. However, further evaluation in practical settings is required to assess our approach when applied to existing R\&S specifications.

Along with previous studies, our approach emphasizes that GDPR is concerned with both \emph{What} and \emph{How} of software engineering. The comparison of our approach to other existing approaches reveals the complexity of the relationships between requirements and system components that GDPR describes and that should be captured across R\&S specifications (e.g., relationships between requirements and concrete system components).

Our results support the existing assumptions about only conditional delineation between software requirements and architecture~\cite{de2009similarity}. The text of GDPR indeed addresses simultaneously problem and solution spaces in their conjunction. This makes it essential to model the close relationships between requirements and (early) system components to achieve PbD as prescribed by GDPR. However, we consider it to be essential to delineate R\&S specifications to differentiate between the instances of concepts requiring further interpretation (requirements specification) and instances of concepts not requiring interpretation (system specification).

During the interviews, in many cases, different interviewees mentioned the same specification goal in relation to different specification objectives. For example, the purpose of reacting to changes was mentioned in relation to SO5, SO2, and SO1. This reflects the complexity of how specification objectives (e.g., traceability), contribute to different goals (e.g., reacting to changes). Further, in-depth research into these relationships remains the subject of our future research. The results of our experiment identified some of the complexities of addressing legal knowledge which are relevant both for manual and automated processing of regulations (e.g., identification of semantic synonyms also constitutes a challenge in large language mode-based text processing~\cite{hayashi2025evaluating}). We plan to explore these challenges in our future work.

In our future work, we plan to address the operationalization of the proposed approach to enable its application in industrial settings. Building on the results of the interviews, we aim to explore operationalization using templates or tools that would be suitable across different organizational settings and engineering models. We also plan to apply the approach to auxiliary regulatory sources that complement GDPR, as well as to other regulations that demand fulfillment of compliance by design. Furthermore, we aim to conduct a case-based evaluation of the operationalized approach. Such an evaluation will also include empirical investigation of the capacity of the approach to support the achievement of the specification objectives that practitioners consider most important. Finally, we will also investigate whether certain specification objectives hold greater importance for specific software engineering roles compared to others.

We suggest that the following practical implications emerging from our study are important for practitioners:
\begin{itemize}[nosep]
    \item It is essential to further raise awareness and facilitate experience sharing on the topics of GDPR compliance and PbD, as practitioners struggle to systematically reason about PbD and GDPR compliance.
    \item Specification objectives (e.g., traceability) and goals (e.g., risk management) can serve as guidance for the specification process to ensure effective implementation of privacy by design.
    \item Conjoint consideration of requirements and system specifications is essential to capture the GDPR intentions as a basis of privacy by design.
    \item Traceability is essential in privacy by design implementation; however, it should be supported by (1) availability of legal domain knowledge in a usable form, (2) transparency and understandability of specifications enabling communication of different roles involved.
    \item Requirements engineering methods capturing legal domain knowledge in a structured way can facilitate requirements and system (architecture) specification.
\end{itemize}

\section{Threats to validity}\label{sec:threatsValidity}
In this section, we summarize the main threats to validity and the strategies we applied to mitigate them. We use the approach of Wohlin et al.~\cite{wohlin2012experimentation} and report conclusion, internal, construct, and external validity threats.

To mitigate threats to the validity of the results, we followed available guidelines for the methods we used in our study. In particular, we followed the guidelines for literature review by Kitchenham~\cite{kitchenham2007guidelines}, guidelines for snowballing by Wohlin~\cite{wohlin2014guidelines}, guidelines for thematic analysis by Braun\&Clarke~\cite{braun2006using}, guidelines for interviews by Linaaker et al.~\cite{linaaker2015guidelines} and Runeson et al.~\cite{runeson2009guidelines}, and guidelines for evaluation of software engineering methods by Kitchenham~\cite{kitchenham1996desmet} and experimentation by Wohlin~\cite{wohlin2012experimentation}.

\subsection{Construct Validity}
\paragraph{Literature Review}
To address the potential threats during the data extraction, the first and the second authors formulated selection criteria that do not require subjective interpretation (e.g., the study contains requirements or evaluation criteria). During the extraction and processing of the extracted data, the first and second authors had meetings to jointly discuss the results. Also, requirements and system specification objectives, synthesized as the result of the literature review, were discussed in a larger circle of authors.

\paragraph{Interviews}
To address the threats of incorrect interpretation of the interview results, we applied interview questions formulated on the basis of the structured Goal Question Metric approach, scoping answers of the interviewees around one of the goals, questions, or metrics relevant to one of the specification objectives. To make sure that we addressed all the relevant R\&S specification concerns, at the end of the interview sessions, we asked interviewees about any other objectives, goals, questions, and metrics that are important to consider, but which were not covered.

\paragraph{Evaluation}
In this study, we aimed to conduct an initial evaluation of our candidate approach; we focused only on a limited scope of regulation specification concepts. To mitigate this threat, we selected the core legal concept relevant to R\&S specification(in particular ``legal object''). This enabled us to focus on the contribution of the underlying conceptual basis of the candidate approach to specification objectives.

We applied two evaluation methods (method screening using description and examples, and experiment) to allow participants to understand the method and its usefulness better. We also asked evaluation participants for any additional comments.  We included a ``3---No opinion'' evaluation option to allow participants to select it when the method screening and the experiment were insufficient to assess the candidate approach.

To minimize the subjective interpretation of each specification objective, we analyzed the evaluation results along with the interview results that allowed tracing to a concrete interpretation of the specification objective by evaluation participants, when required.

\subsection{Internal Validity}
\paragraph{Literature Review}
To address the potential threats to the validity of the literature review results, we documented the process of the literature review. Throughout the whole process, the first two authors actively and regularly discussed any potential challenges. In particular, the first two authors discussed and tested that the selection criteria and the categories of data requiring extraction do not involve any subjective interpretation.

\paragraph{Interviews}
To control for the consistency of the interview results, we used both rating(during the interviews) and ranking(after the interviews) of specification objectives.

We intentionally used different scales for rating(from 5---high importance to 1---low importance) and ranking(from 1---most important to 5---least important). However, after analyzing the results of the first five interviews(I1-I5) we found inconsistencies between rating and ranking that could indicate that some of the interviewees applied the same assessment approach(with 5 as the highest importance) both in rating and ranking. We contacted the interviewees and pointed out this discrepancy. Among the five interviewees, three(I2, I4, I5) changed their ranking, and two others(I1 and I3) confirmed the rating and ranking they assigned. As a result, inconsistency between rating and ranking persisted for I3 who assigned SO3(separation of concerns) the lowest importance (1) and ranked it as the most important (1). Despite this inconsistency, we report the results as confirmed by the I3 and do not consider that this discrepancy has a significant impact on the overall results of the interviews. During the subsequent interviews, we emphasized the difference between the assessment approaches for rating and ranking, and we have not observed any further discrepancies.

Furthermore, to check the correctness of the rating and ranking of specification goals and goals, questions, and metrics identified by the interviewees, we executed member checking and sent interviewees follow-up emails with a summary of the interview results and asked them to contact us in case of any corrections or changes were required.

\paragraph{Evaluation}
To address the threats related to the instrumentation during the evaluation, we asked participants about the clarity of the candidate approach description, examples, and tasks given in the experiment before they started executing the tasks. After the evaluation participants completed the annotation tasks, we demonstrated them with the annotations and instances of the model produced by the first author of this study, and only then requested to evaluate the candidate approach. We have not observed that the demonstration of the ground truth had an impact on the candidate approach evaluation.

\subsection{External Validity}
There was a wide range of roles involved in the interviews and the experiment. We mitigated this threat by scoping our call for participation according to participants' engagement in software requirements and system specifications for GDPR compliance, and the availability of insights into both GDPR compliance and software technologies.

Our evaluation focused on excerpts from four GDPR Articles (4,6,13,15). To mitigate this threat we applied the approach to the whole GDPR and compared the derived requirements and system components suggested in other studies. We also selected the articles that are central to GDPR and are often used as examples in other studies.

\subsection{Conclusion Validity}
For the analysis of the results of our experiment, we followed an interpretive approach; however, we applied some quantitative measurements. Answering RQ8 on to what degree practitioners can use our approach, we used quantified measurement of the annotation conducted by participants and components identified. The application of such quantified measurement remains limited in cases when annotation of the correct part of the text was not conducted completely or when the correct part of the text was incorrectly annotated by a few other concepts. To mitigate this threat to validity, we disclose the annotations done by participants in the open dataset.

We used a subjective measurement of the usefulness of our candidate approach for the achievement of one of the specification objectives. To address the challenge of measurement subjectivity, we selected a very specific scope for the usefulness assessment---specification objectives related to requirements and system specification for GDPR compliance.

To address the potential threats to the reliability of treatment implementation, we used similar handouts or templates for the method screening and experiment in all of the evaluation sessions. We avoided providing any comments or discussing anything related to tasks during the evaluation sessions.

Our conclusions are based on the evaluation of the instantiation of the content model with only some of the relevant legal concepts. However, we consider it to be relevant for the conceptual validation and creation of the foundation for evaluation in industrial settings. As our results show, these selected concepts were sufficient for the evaluation participants to assess our candidate approach. Among 45 evaluations given, only 4 were ``3---No opinion''. Evaluation in industrial settings involving all the required concepts is in the scope of our future work.

\section{Conclusion}\label{sec:conclusion}
Implementation of privacy by design according to GDPR and compliance by design with other regulations is becoming essential in software engineering. However, the complexity of GDPR and legal knowledge, considering both problem and solution space with their multiple interdependencies, creates challenges to the implementation of GDPR compliance and privacy by design. It is essential to address the existing gap in practice for conjoint requirements and system specification with determined specification objectives.

Our research shows that specification objectives of reflecting legal knowledge, specification transparency, and traceability are essential in the implementation of privacy by design. Future research on privacy by design should take into account the demand to sufficiently reflect legal knowledge and interact with multiple stakeholders. It is also important to research the impact that legal knowledge reflection and the involvement of different stakeholders have on the implementation of traceability in terms of understandability and granularity of the traced information. We suggest that further empirical research on specification objectives for privacy by design is required, taking into account that specification practices remain insufficiently established.

Specification objectives for privacy by design should be achieved both with requirements and system specifications. Hence, conjoint consideration of both specifications is essential. This is also emphasized by specification goals (e.g., compliance feasibility) that interviewees identified and which are achieved throughout both requirements and system specifications. We suggest that future research could focus on better understanding the eventual goals to which specification methods should contribute.

The results of our evaluation confirm that the premise of capturing legal domain knowledge is important to achieve some of the specification objectives. It contributes to achieving transparency of specifications and potentially to traceability and facilitation of system flexibility. The application of our candidate approach, closely capturing legal concepts, also results in closely integrated requirements and system specifications. The results of our experiment suggest that it is challenging for practitioners to effectively apply our conceptual model to the text of GDPR and build corresponding specification content models. Our future research will be focused on the identification of effective approaches to operationalize the model in industrial settings.

\paragraph{Data Availability Statement}
An open dataset for this study is available on \href{https://zenodo.org/records/15565632}{Zenodo (DOI: 10.5281/zenodo.15565632)} and includes the following data:
\begin{itemize}[nosep]
    \item results of the literature search and selection;
    \item mapping of the data extracted from the literature to synthesized R\&S specification objectives;
    \item questionnaire used during the interviews;
    \item specification goals identified by interviewees;
    \item materials used in the candidate approach evaluation;
    \item results of the experiment for each evaluation participant.
\end{itemize}

\paragraph{Acknowledgements}
The first author expresses his sincere gratitude to all colleagues and practitioners who contributed to this research through their participation, encouragement, and valuable insights into the state of practice. The author extends special thanks to Andreas Essing (IF-Blueprint AG), Daniel Sack (TechTalk), Dr. Rüdiger Lincke (Softwerk AB), Jörn Koch (WPS GmbH), Olha Halytska (LEONET GmbH), Lucas Rangel Meira (neoshare AG) for their thoughtful input and collaboration – your contributions have been truly instrumental in advancing research and practice in privacy engineering.

\footnotesize
\bibliographystyle{elsarticle-num}
\bibliography{bibliography}

\end{document}